\documentclass[
               reprint, twocolumn,
               preprintnumbers,
               amsmath,
               amssymb,
               showpacs,
               aip,jcp,
               superscriptaddress]{revtex4-1}

 \usepackage{ucs}
 \usepackage[utf8x]{inputenc}
 \usepackage{latexsym}
 \usepackage{amsfonts}
 \usepackage{amsmath}
 \usepackage[dvips]{graphicx}
 \usepackage{color}
 \usepackage{bm}
 \usepackage{hyperref}
\hypersetup{
   colorlinks=true,       
   linkcolor=magenta,          
   citecolor=magenta,        
   filecolor=magenta,      
   urlcolor=blue           
}

\begin{document}

\author{Sudip Sasmal}
\email[e-mail: ]{sudip.sasmal@pci.uni-heidelberg.de}
\affiliation{Theoretische Chemie,
             Physikalisch-Chemisches Institut,
             Universität Heidelberg,
             Im Neuneheimer Feld 229, 69120 Heidelberg, Germany}

\author{Oriol Vendrell}
\email[e-mail: ]{oriol.vendrell@pci.uni-heidelberg.de}
\affiliation{Theoretische Chemie,
             Physikalisch-Chemisches Institut,
             Universität Heidelberg,
             Im Neuneheimer Feld 229, 69120 Heidelberg, Germany}
\affiliation{Interdisciplinary Center for Scientific Computing,
             Universität Heidelberg,
             Im Neuneheimer Feld 205, 69120 Heidelberg, Germany}




\title{Non-adiabatic quantum dynamics without potential energy surfaces based on
second-quantized electrons: application within the framework of the MCTDH method}

\date{\today}

\begin{abstract}
    A first principles quantum formalism to describe the non-adiabatic dynamics of
    electrons and nuclei based on a second quantization representation (SQR) of
    the electronic motion combined with the usual representation of the nuclear
    coordinates is introduced.
    This procedure circumvents the introduction of potential energy surfaces and
    non-adiabatic couplings, providing an alternative to the Born-Oppenheimer
    approximation.
    An important feature of the molecular Hamiltonian in the mixed first
    quantized representation for the nuclei and the SQR representation for the
    electrons is that all degrees of freedom, nuclear positions and electronic
    occupations, are distinguishable.
    This makes the approach compatible with various tensor decomposition
    \emph{ansätze} for the propagation of the nuclear-electronic wavefunction.
    Here, we describe the application of this formalism within the
    multi-configuration time-dependent Hartree (MCTDH) framework and its
    multilayer generalization, corresponding to Tucker and hierarchical Tucker
    tensor decompositions of the wavefunction, respectively.
    The approach is applied to the calculation of the photodissociation
    cross-section of the HeH$^+$ molecule under extreme ultraviolet irradiation,
    which features non-adiabatic effects and quantum interferences between the
    two possible fragmentation channels, He+H$^+$ and He$^+$+H. These
    calculations are compared with the usual description based on \emph{ab
    initio} potential energy surfaces and non-adiabatic coupling matrix
    elements, which fully agree.
    The proof-of-principle calculations serve to illustrate the advantages and drawbacks
    of this formalism, which are discussed in detail, as well as possible ways
    to overcome them. We close with an outlook of possible application domains
    where the formalism might outperform the usual approach, for example in
    situations that combine a strong static correlation of the electrons with
    non-adiabatic electronic-nuclear effects.
\end{abstract}


\maketitle

\section{Introduction} \label{sec:introduction}
    %
    The Born-Oppenheimer approximation (BOA)~\cite{bor27:457, Born-Crystal} and
    its group BOA generalization~\cite{Born-Crystal, pac93:293, wor04:127}
    represent the cornerstone on which the traditional description of the
    structure of molecules and of the coupled electron-nuclear motion is based.
    They separate the electronic and nuclear contributions to the molecular
    wavefunction
    through the introduction of electronic states parametrized by the nuclear
    positions,
    thus resulting in chemical structures and reactions determined by the
    position and displacements of nuclei over the potential energy surface(s)
    (PES) provided by the electrons.
    The BOA is an excellent description of ground-state chemistry but its
    failure is quite common in the photophysics and photochemistry of molecular
    systems~\cite{Tru03:32501,dom04}.
    It breaks down when the energetic gap between the relevant PESs becomes of
    the order of the spacing between vibrational states,
    i.e. in the vicinity of avoided crossings and conical intersections where two or more
    PES interact strongly via vibronic coupling~\cite{azu77:315,kop84:59,
    Tru03:32501, Bersuker2011, dom04, conical2011}.
    The group BOA (GBOA) is the canonical strategy to deal with these
    situations~\cite{pac93:293, wor04:127}. A group of electronic states assumed
    uncoupled with the rest is introduced and the non-adiabatic couplings
    within the group are considered in either adiabatic or diabatic
    representations, this choice being often a matter of practical convenience.
    Multiple theoretical methods and strategies exist to both: computing the
    non-adiabatic couplings (NAC) between electronic states of the
    group~\cite{lis04:7322,Li14:244105,alg14:7140,rya15:4200}
    and obtaining a suitable \mbox{(quasi-)}diabatic representation of the coupled
    nuclear and electronic
    Hamiltonian~\cite{kop84:59,pac88:7367,yar98:8073,koe01:2377,nak02:5576,jor12:084304,yan13:84}.

    %
    Although the GBOA is possibly the most practical strategy to deal with
    chemical problems in which only a handful of electronic states play a
    significant role,
    the calculation of NACs and quasi-diabatic states becomes a formidable task
    for situations involving a large number of energetically close-lying
    PES, for example when metal
    centers~\cite{vee10:67401,gra10:4550} and highly excited
    electronic states are involved~\cite{Li14:457,Li16:043203}.
    These situations arise as well in energy and charge transfer scenarios,
    where the
    electronic and nuclear dynamics become inextricable
    and
    which are often approached via model
    Hamiltonians~\cite{ren01:137,tam15:107401,red19:044307,May2004,sch15:1}.
    %
    %
    Nonetheless, GBOA non-adiabatic dynamics, both fully quantum-mechanical
    and mixed quantum-classical,
    and involving a large number of electronic states, are still feasible under
    certain
    approximations~\cite{lis11:033408,jor12:084304,Li16:043203}
    and there is much interest in the
    further development of approaches to obtain NACs in dense
    electronic manifolds~\cite{sub08:244101,sub16:387}.
    However, it is interesting to consider the fact that the specific electronic
    state occupied by the system at every time during the relevant dynamics is
    often not an experimental observable that needs to be determined, but rather
    an auxiliary device along the calculation of, e.g., rates, time-scales and
    final outcomes of chemical processes.

    %
    In this work, we propose an alternative \emph{ab initio} approach that
    circumvents the introduction of a group of electronic states and the
    calculation of the corresponding PES and NACs.
    It is based on the second quantization representation (SQR) of the
    electronic subsystem~\cite{Fetter2003}
    where the electronic Hamiltonian acts on the Fock space
    spanned by occupation-number states constructed from a set of properly
    diabatized single-particle electronic orbitals.
    In turn,
    the nuclear coordinates are treated within the usual first quantization
    framework.
    In this nuclear-SQR (N-SQR) formalism, the non-adiabatic effects are
    described by the evolution of the nuclear amplitudes coupled to the
    dynamics of the orbital occupations of the underlying single-particle
    basis for the electrons.
    The combination of a first quantized description of vibrational
    (or phononic) degrees of freedom and a second quantized description of the
    electrons is not new. For example, the combination of Hubbard's~\cite{hub63:238}
    and Holstein's~\cite{hol59:325} models to describe correlated electronic and lattice
    motions has a history of more than 40 years in the field of solid-state
    physics~\cite{cor74:997}.
    Taking this route to describe the coupled non-adiabatic dynamics of molecular systems
    from an \emph{ab initio} perspective has, to the best of knowledge, not been
    attempted yet.
    %

    An important feature of the N-SQR formalism is that all degrees of freedom
    of the molecular Hamiltonian, nuclear positions and electronic occupations,
    are distinguishable, based on the fact that in a SQR
    representation the particles' indistinguishability is determined by the
    commutation relations of the corresponding creation and annihilation
    operators, and not by the symmetry of the wavefunction~\cite{Fetter2003}.
    This feature permits the straightforward application of low-rank tensor
    approximations to the full nuclear-electronic wavefunction, for example
    Tucker and hierarchical Tucker decompositions, or matrix-product states,
    which is not possible if the electronic part of the wavefunction is to be
    described by a pair-wise antisymmetric tensor~\cite{lat00:1253,cer20:591}.
    %
    %
    %


    Here, we focus on the representation and the propagation of the N-SQR
    wavefunction within the framework of the
    multiconfiguration time-dependent Hartree MCTDH~\cite{mey90:73, man92:3199,
    bec00:1, mey03:251,mey09:book,mey12:351} method and machinery for
    second quantized Hamiltonians
    (MCTDH-SQR)~\cite{wan09:024114,wan13:134704,man17:064117,wei20:034101} (while
    including first-quantized nuclear degrees of freedom) and note that other
    methodological alternatives might be suitable as well.
    Combining first quantized vibrations and second
    quantized electrons has been successfully achieved by Wang and Thoss within the context of
    the MCTDH-SQR approach in applications to model Hamiltonians~\cite{wan09:024114,wan13:134704}.
    The MCTDH \emph{ansatz} corresponds to an optimal time-dependent Tucker
    decomposition of the multi-dimensional wavefunction; its multilayer
    (ML-MCTDH)~\cite{wan03:1289, man08:164116, Ven11:44135} generalization
    corresponds to a hierarchical Tucker decomposition and is therefore also
    N-SQR compatible. Both the MCTDH and ML-MCTDH \emph{ansätze} do not consider
    any particular symmetry between the underlying degrees of freedom.
    %

    %
    Circumventing the (G)BOA to describe the coupled electronic and nuclear
    dynamics of molecules is a problem of practical and fundamental
    significance that has been addressed by multiple researchers and from
    diverse perspectives~\cite{%
        hax11:63416,%
        ulu12:054112,%
        muo18:184105,%
        aeb19:023406}.
    A non-exhaustive overview of such efforts to follows.
    One can construct the total multiconfiguration wavefunction using products
    of nuclear wavefunctions of electronic Slater
    determinants~\cite{hax11:63416,ulu12:054112}.
    This corresponds, in short, to extensions of the MCTDH approach for fermions
    (MCTDH-F)~\cite{cai05:12712,alo07:154103,hoc11:084106,sat13:023402,lod20:011001}
    to a direct product of time-dependent Slater determinants and
    an orthonormal nuclear functions.
    A drawback of such approach is that the primitive basis functions of
    nuclei and electrons are, strictly speaking, independent of each other. This
    means, the strong but trivial correlation between the positions of the
    nuclei and electrons due to their Coulomb attraction is carried into the
    nuclear-electronic wavefunction and needs to be propagated.
    Haxton and McCurdy resolved this problem for the special case of
    diatomic molecules by introducing prolate spheroidal
    coordinates for the electronic wavefunction~\cite{hax11:63416}.
    Ulusoy and Nest introduced a set of atomic orbitals on ``ghost'' centers
    along the internuclear axis of diatomic systems, thus increasing the
    effective size of the electronic basis but avoiding the re-evaluation of
    matrix elements at each nuclear position~\cite{ulu12:054112}.
    Due to the description of the electrons in an optimal basis of time-dependent
    orbitals inherited from the variational MCTDH-F framework, these treatments
    have been successful at describing small molecular systems in intense laser
    fields~\cite{hax11:63416,ulu12:054112,sat13:023402,aeb19:023406}.
    A limitation of such descriptions, also inherited from MCTDH-F, is that the
    number of configurations of the electronic subsystem increases
    combinatorially with the number of electrons and electronic single-particle
    functions, thus hindering their application beyond small molecular systems.
    Active-space generalizations of the MCTDH-F method can help in mitigating
    this scaling~\cite{lod20:011001}.

    Reiher and co-workers introduced explicitly correlated basis functions
    between electrons and nuclei; functions that depend on the
    electronic-nuclear distances. With this treatment they computed
    vibrational-electronic eigenstates of small molecules in a
    pre-Born-Oppenheimer fashion and without relying on the concept of molecular
    structures, i.e. without introducing a nuclear framework \emph{a priori} and
    thus challenging the traditional concept of molecular
    structure~\cite{muo18:184105}.
    Finally, approaches based on an exact, single-product factorization of the
    nuclear-electronic wavefunction~\cite{abe10:123002,ced13:224110,ago13:3625}
    or using conditional electronic
    wavefunctions~\cite{alb14:083003,alb19:023803} also circumvent the BOA.
    These types of approaches lead to the picture of a nuclear wavefunction
    evolving on a time-dependent potential whose equation of motion requires
    knowledge of the nuclear amplitudes. This results in a complex set of
    coupled equations that have been mostly deployed on model
    Hamiltonians~\cite{ago13:3625,alb19:023803}, but which have also been the
    basis of improved mixed quantum-classical approaches to non-adiabatic
    dynamics~\cite{min17:3048}.
    Closing this overview, it is worth mentioning as well that Reiher has
    recently proposed a method based on quantum dynamics for electrons and
    nuclei based on matrix product states that bears formal resemblances to
    N-SQR~\cite{bai19:3481}.  Nonetheless, their approach represents the
    wavefunction of a system consisting of several nuclear degrees of freedom
    and a discrete set of electronic states, i.e. in the GBOA framework, using
    the time-dependent density matrix renormalization group (TD-DMRG)
    \emph{ansatz}. In this respect, the method is in its aim similar to
    applications of the ML-MCTDH algorithm to non-adiabatic dynamics
    situations~\cite{Ven11:44135,men13:14313}.

    The paper is organized as follows.
    Sections~\ref{sec:theory:mctdh} and \ref{sec:theory:sqr} shortly review the
    fundamental aspects of MCTDH and of its application to indistinguishable
    fermions in a SQR, respectively.
    Section~\ref{sec:theory:n-sqr} introduces the N-SQR Hamiltonian whereas
    Sec.~\ref{sec:theory:impl} discusses practical and implementation details
    related to using the N-SQR approach in the framework of MCTDH.
    Section~\ref{sec:computation} describes the computational details of the
    proof-of-concept application to the photodissociation cross section of the
    HeH$^+$ molecule, on which the N-SQR approach is compared to GBOA
    calculations.
    Section~\ref{sec:results} presents and discusses the results and finally a
    summary and conclusions including an outlook is provided in
    Sec.~\ref{sec:conclusions}.

\section{Theory}            \label{sec:theory}
\subsection{MCTDH formalism}  \label{sec:theory:mctdh}
In the MCTDH formalism \cite{mey90:73, man92:3199}, the $f$-dimensional wavefunction
$\Psi(x_1,x_2,\dots,x_f,t)$ is expanded in an orthonormal basis of time-dependent single
 particle functions (SPFs)
\begin{widetext}
\begin{eqnarray}
 \Psi(x_1,\dots,x_f,t) =  \sum_{j_1=1}^{l_1}\dots\sum_{j_f=1}^{l_f} A_{j_1\dots
    j_f}(t)\prod_{\kappa=1}^{f}\phi_{j_k}^{(\kappa)}(x_\kappa,t)
 = \sum_{J}A_J(t)\Phi_J(t)
\end{eqnarray}
\end{widetext}
where the usual nomenclature of the MCTDH literature is used~\cite{bec00:1}
except for the number of SPFs in each DOF $l_\kappa$, which is usually labeled by
$n_\kappa$ and which we reserve for later use.
$A_J(t)$ is the time-dependent expansion coefficient of the $J$-th configuration marked with
multi-index $J$, and $\Phi_J(t)$ is the $J$-th time-dependent Hartree product that is formed by
a direct product of SPFs for each degree of freedom (DOF).
The SPFs ($\phi_{j_\kappa}^{(\kappa)}(x_\kappa,t)$) are typically represented in a time-independent (primitive) basis
\begin{eqnarray}
    \label{eq:spfs}
    \phi_{j_\kappa}^{(\kappa)}(x_\kappa,t) = \sum_{i_\kappa}^{N_\kappa}
    c_{i_\kappa, j_\kappa}^{(\kappa)}(t)\chi_{i_\kappa}^{(\kappa)}(x_\kappa)
\end{eqnarray}
where $c_{i_\kappa, j_\kappa}^{(\kappa)}(t)$ is a time-dependent expansion
coefficient of the $j$-th SPF of the $\kappa$-th DOF.
For convenience, these primitive basis functions
($\chi_{i_\kappa}^{(\kappa)}(x_\kappa)$) are often chosen as a discrete variable
representation (DVR), thus greatly simplifying the evaluation of the matrix
elements of the potential energy operator\cite{bec00:1}.

Therefore, one can view the MCTDH \emph{ansatz} as a two layer wavefunction
where, in the bottom layer, the $c_{i_\kappa, j_\kappa}^{(\kappa)}(t)$
coefficients represent optimally evolving time-dependent SPFs along each
coordinate, whereas the upper layer of $A_J(t)$ coefficients represents the
total multi-dimensional wavefunction in the direct products basis of the
time-dependent SPFs.
In MCTDH, one often combines the $f$ physical coordinates ($x_1,x_2,\dots,x_f$)
into $d$ groups of logical multidimensional coordinates ($q_1,q_2,\dots,q_d$)
\cite{mey03:251}.
This has no effect in the equations of motion (EOM) but is an important element
to balance the cost of the propagation between the A-vector and the SPFs.

Finally, the EOM for the MCTDH ansatz are obtained using the time
dependent Dirac-Frenkel variational principle~\cite{bro88:547}
\begin{align}
 \langle \partial \Psi | H-i \frac{\partial}{\partial t} | \Psi \rangle = 0 .
\end{align}
With the constrains $\langle \phi_j^{(\kappa)} | \dot{\phi}_i^{(\kappa)} \rangle = 0$, the
EOM for the MCTDH ansatz read~\cite{bec00:1}
\begin{align}
    i\dot{A}_J & = \sum_L \langle \Phi_J | H | \Phi_L \rangle \dot{A}_L \\
    i\dot{\boldsymbol{\phi}}^{(\kappa)} & =
    (\boldsymbol{1}-\boldsymbol{P}^{(\kappa)})(\boldsymbol{\rho}^{(\kappa)})^{-1}
    \langle \boldsymbol{H} \rangle^{(\kappa)} \boldsymbol{\phi}^{(\kappa)}
\end{align}
where the vector notation $\boldsymbol{\phi}^{(\kappa)} = (\phi_{1}^{(\kappa)},
\phi_{2}^{(\kappa)}, \dots , \phi_{l_\kappa}^{(\kappa)})^T$ is used.
$\boldsymbol{P}^{(\kappa)}$ is the projector on the space spanned by the SPFs of
the $\kappa$-th DOF and $\boldsymbol{\rho}^{(\kappa)}$ and $\langle
\boldsymbol{H} \rangle^{(\kappa)}$ are the density matrix and the mean field of
$\kappa$-th DOF \cite{bec00:1}.
The efficiency gain in  MCTDH compared to the standard method (propagating
directly on the primitive basis) arises from the fact that the number optimal of
time-dependent configurations that need to be propagated to describe the
correlation of the system is usually much smaller than the number of primitive
configurations ($\prod_{\kappa=1}^{f} l_\kappa$ compared to
$\prod_{\kappa=1}^{f} N_\kappa$).
It is, of course, possible to identify counterexamples for very highly
correlated systems~\cite{hin16:012009}. In such cases, the number of
time-dependent configurations needed to achieve convergence with MCTDH
approaches the number of primitive configurations and the overhead of the
algorithm renders it less efficient. 
Mode combination \cite{mey03:251} and especially the multilayer (ML) approach
\cite{wan03:1289, man08:164116, Ven11:44135, cao13:134103}, which is implemented
in the Heidelberg MCTDH package \cite{Ven11:44135, mctdh:MLpackage}, can boost
the efficiency further.  This has allowed to describe systems with hundreds and
up to thousands of DOF for some model Hamiltonians \cite{cra07:144503,
wan08:139, Ven11:44135, shi17:184001}.


In the MCTDH formalism introduced above, the coordinates $x_\kappa$
correspond to distinguishable DOF. Hence, the \emph{ansatz} can
be regarded as a tensor contraction in Tucker format
\begin{align}
    \label{eq:tucker_mctdh}
    C_{i_1,\ldots,i_f}(t) =
        \sum_{j_1,\ldots,j_f}^{l_1,\ldots,l_f} A_{j_1,\ldots,j_f}(t)
        \prod_{\kappa=1}^f c_{i_\kappa,j_\kappa}^{(\kappa)}(t)
\end{align}
of the expansion coefficients
of the primitive basis functions $C_{i_1,\ldots,i_f}(t)$.
The coefficients $c_{i_\kappa, j_\kappa}^{(\kappa)j}(t)$ can be arranged as
$l_\kappa\times N_\kappa$
matrices while the $A$-vector is the core tensor of the contraction. Hence, its
rank is smaller than the rank of the exact (within the primitive basis)
$C$-vector.  In mode combination~\cite{bec00:1},
the mode tensors $c_{i_\kappa,j_\kappa}^{(\kappa)}$
are chosen of a larger
order than two by turning the $i_\kappa$ index into a multi-index.
Correspondingly, the core tensor rank is further reduced.
Once the mode tensors have too many terms, a Tucker decomposition can be applied
to them, which results in the hierarchical Tucker format of
multilayer-(ML-)MCTDH. The key requirement for this hierachical construct to be
possible is that the primitive indices $\{i_1,\ldots,i_f\}$ refer to a direct
product basis of distinguishable DOF.
%

It is still possible, although inefficient, to describe
indistinguishable DOF within the original MCTDH (not ML) framework.  For this,
the SPFs must be restricted to one single set for all particles and the
propagation must be started with an (anti)symmetric $A$-vector. Early
applications of MCTDH to the field of cold Bosons relied on this
strategy~\cite{zoe06:053612, zoe08:013629}.
Clearly, this case is not amenable to a \emph{hierarchical} Tucker decomposition
because the entries of the $A$-vector in Eq.~(\ref{eq:tucker_mctdh}) are related by
symmetry upon permutation of two indices: the grouping of particle DOFs in
logical coordinates is not possible.
However, the original MCTDH theory has been extended by introducing permanents
for Bosons~\cite{alo07:154103,alo08:33613} and Slater determinants for
Fermions~\cite{cai05:12712,alo07:154103,hoc11:084106,sat13:023402,lod20:011001},
which avoids the presence of redundant (exchange symmetry-related) A-vector
entries in the wavefunction. These approaches are made efficient by using
standard rules to calculate the matrix elements of the (anti-)symmetric
configurations, e.g. Slater-Condon rules for Fermions.
%
%
%

\subsection{MCTDH-SQR} \label{sec:theory:sqr}


\subsubsection{General aspects}

A fundamentally different alternative to describe systems of indistinguishable
particles is to use a second quantization representation (SQR). This approach
was described and applied for the first time by Thoss and Wang in the context of
MCTDH and termed MCTDH in SQR (MCTDH-SQR)~\cite{wan09:024114}.
The underlying idea is quite general and widely used besides the MCTDH
context. Here we describe the general features of this formulation for
the sake of completeness before describing the particular aspects of our
implementation.

In SQR, the state of the system can be described in occupation number
formalism as a superposition of Fock states
\begin{eqnarray}
 \label{eq:occnum}
 | \boldsymbol{n} \rangle = |n_1, n_2, \dots , n_M \rangle,
\end{eqnarray}
where $n_\kappa$ represents the occupation of the $\kappa$-th single particle state (SPS,
to be differentiated from the MCTDH SPF introduced above) and $M$ corresponds to the
total number of SPS.  $n_\kappa$ can be equal
to $0$ or any positive integer for Bosons and either $0$ or $1$ for Fermions.
The indices of the primitive tensor of coefficients in
\begin{align}
    \label{eq:tucker_sqr}
    |\Psi(t)\rangle = \sum_{n_1,\dots,n_M}C_{n_1,\ldots,n_M}(t)
                   |n_1, n_2, \dots , n_M \rangle
\end{align}
refer now to an underlying set of distinguishable degrees of freedom, the
occupations of each SPS of the many-body system.
These Fock space states can be written as well as the direct product of the occupation
number states of the $M$ Fock subspaces, one for each SPS
\begin{eqnarray}
 \label{eq:occnum_hartree}
 | \boldsymbol{n} \rangle = | {n_1} \rangle \otimes |
    {n_2} \rangle \otimes \dots \otimes | {n_M} \rangle,
\end{eqnarray}
where we simply are acknowledging the fact that the coefficients in the
wavefunction~(\ref{eq:tucker_sqr}) can be regarded as an $M$-dimensional tensor.
Because of the distinguishability of the degrees of freedom,
a hierarchical Tucker decomposition of the primitive tensor is
possible and therefore the original MCTDH formalism (or its multilayer
generalisation) is, in principle, applicable without further modification.


What remains to be discussed is how the Hamiltonian of the many-body system acts
on the wavefunction, and this is different for bosons and for fermions.
MCTDH is most efficient when the operator takes the form of a sum of product
terms, each acting on a degree of freedom.
The SQR operator for a many-body system of bosons or fermions (with up to
two-body interactions) reads
\begin{eqnarray}
    \label{eq:sqHam}
 \hat{H} &=& \sum_{ij} h_{_{ij}} \hat{a}_i^{\dagger}\hat{a}_{_j} + \frac{1}{2}
    \sum_{ijkl}
    v_{_{ijkl}}\hat{a}_i^{\dagger}\hat{a}_j^{\dagger}\hat{a}_{_l}\hat{a}_{_k},
\end{eqnarray}
where the $\hat{a}_i^{(\dagger)}$ correspond to the annihilation (creation)
operators fulfilling the corresponding commutation relations and
\begin{eqnarray}
  h_{_{ij}} &=& \langle \phi_i(1)| -\frac{1}{2}\nabla_1^2 - \sum_{A=1}^M \frac{Z_A}{r_{_{1A}}} | \phi_j(1) \rangle \\
    v_{_{ijkl}} &=& \langle \phi_i(1) \phi_j(2) | v(1,2) | \phi_k(1) \phi_l(2) \rangle
\end{eqnarray}
are the one- and two-body integrals, respectively, involving the
SPSs $\phi_i(1)$.
Although our primary interest is in the fermionic (electronic) case,
we describe first the treatment of bosons, which is conceptually simpler.

The creation and annihilation operators for bosons fulfill the commutation
relations
 \begin{align}
     \label{eq:antic1}
     [\hat{a}_i, \hat{a}_j^{\dagger} ] & =
    \hat{a}_i\hat{a}_j^{\dagger} - \hat{a}_j^{\dagger}\hat{a}_i
    = \delta_{ij} \\
     \label{eq:antic2}
     [\hat{a}_i^{\dagger}, \hat{a}_j^{\dagger} ] & = [ \hat{a}_i, \hat{a}_j ] = 0,
 \end{align}
which imply that the creation or annihilation of a particle in
a specific SPS does not carry with it a multiplicative phase factor depending on the
occupation of the other SPSs~\cite{Fetter2003}.
Additionally, for $j=i$ the relations~(\ref{eq:antic1}, \ref{eq:antic2})
coincide with the
ladder operators of a harmonic oscillator. Thus, as is well known,
the Hamiltonian of the many-body bosonic system
is isomorph with the Hamiltonian of a set of coupled harmonic oscillators,
where each of them represents the particle occupation of a SPS.
Hence, the Hamiltonian~(\ref{eq:sqHam}) is, for the bosonic case, of product
form.
In practical applications, e.g. within MCTDH, one can either directly use the
matrix representation of the ladder (creation/annihilation) operators, or introduce
their standard form as a function of position and momentum operators.
Multiple applications of the MCTDH method are found in the literature, where
coupled oscillators are involved. In particular, the ML-MCTDH approach has been
used to describe about 1500 coupled oscillators in the Henon-Heiles
model~\cite{Ven11:44135},
and other works considering even larger numbers of degrees of
freedom can be found~\cite{cra07:144503, wan08:139}.

For the case of fermions, the commutation relations are
\begin{align}
 \label{eq:antic3}
 \{\hat{a}_i, \hat{a}_j^{\dagger} \} =
    \hat{a}_i\hat{a}_j^{\dagger} + \hat{a}_j^{\dagger}\hat{a}_i 
    = \delta_{ij} \\
 \label{eq:antic4}
  \{\hat{a}_i^{\dagger}, \hat{a}_j^{\dagger} \} = \{\hat{a}_i, \hat{a}_j \} = 0.
\end{align}
These have as a consequence that the operator $\hat{a}_s$
acting on a Fock-space
\emph{ket}
\begin{align}
    \label{eq:opfermi}
    \hat{a}_s |n_1, n_2, \dots , n_M \rangle  =
            \hat{a}_s (\hat{a}_1^\dagger)^{n_1}
                      (\hat{a}_2^\dagger)^{n_2} \cdots
                      (\hat{a}_M^\dagger)^{n_M}
                      |\mathrm{vac}\rangle \\ \nonumber
         =
            (-1)^{S_s} (\hat{a}_1^\dagger)^{n_1}
                       (\hat{a}_2^\dagger)^{n_2}    \cdots
                       (\hat{a}_s\hat{a}_s^\dagger) \cdots
                      (\hat{a}_M^\dagger)^{n_M}
                       |\mathrm{vac}\rangle
\end{align}
accumulates a phase factor $S_s = \sum_{k=1}^{s-1} n_k$ that depends
on the occupation of all SPS (spin-orbitals
for the electronic case) before the $s$-th position, and the same is true for
$\hat{a}_s^\dagger$~\cite{Fetter2003}.
This phase complicates the application of Hamiltonian~(\ref{eq:sqHam}) to the
wavefunction. Clearly, the operators $\hat{a}_s^{(\dagger)}$ operate beyond
their index $s$ and therefore are, in general, not of product form with respect
to the primitive degrees of freedom or combined modes used in the MCTDH
\emph{Ansatz}.

One possible solution, the one we adopt here and also the one proposed by Wang
and Thoss~\cite{wan09:024114},
is to map the fermionic SQR Hamiltonian onto an equivalent spin
Hamiltonian.
Formally, this is done by mapping fermionic DOF onto spin DOF using
the (inverse) Jordan–Wigner
transformation~\cite{jor28:631}, according to . i.e.,
 \begin{eqnarray}
 \hat{a}_i^{\dagger} =
     \exp\left( +i \pi \sum_{k=1}^{i-1} \hat{\sigma}_k^{+}\hat{\sigma}_k^{-}  \right)
     \hat{\sigma}_i^{+} \nonumber\\
 \hat{a}_i = 
     \exp\left( -i \pi \sum_{k=1}^{i-1} \hat{\sigma}_k^{+}\hat{\sigma}_k^{-}  \right)
     \hat{\sigma}_i^{-}
 \label{eq:jorwig_1}
 \end{eqnarray}
and
\begin{eqnarray}
    \hat{\sigma}_k^{+}\hat{\sigma}_k^{-} = \hat{n}_k = \frac{1}{2} \left( \hat{\sigma}_k^{z}+1 \right)
\end{eqnarray}
where $\hat{\sigma}_i^{+} = \frac{1}{2}\left( \hat{\sigma}_i^{x}+i\hat{\sigma}_i^{y} \right)$,
$\hat{\sigma}_i^{-} = \frac{1}{2}\left( \hat{\sigma}_i^{x}-i\hat{\sigma}_i^{y} \right)$, and
$\hat{\sigma}_k^{z}$ are the standard spin ladder operators with Pauli matrices
$\bm{\sigma}^{x}$, $\bm{\sigma}^{y}$ and $\bm{\sigma}^{z}$.
Following this prescription,
the creation and annihilation operators (Eq. \ref{eq:jorwig_1}) can be
rewritten as~\cite{wan09:024114}
\begin{eqnarray}
\hat{a}_i^{\dagger} = \prod_{k=1}^{i-1} (-1)^{n_k} \cdot \hat{\sigma}_i^{+} = \prod_{k=1}^{i-1} \hat{S}_k \cdot \hat{\sigma}_i^{+} \nonumber \\
\hat{a}_i = \prod_{k=1}^{i-1} (-1)^{n_k} \cdot \hat{\sigma}_i^{-} = \prod_{k=1}^{i-1} \hat{S}_k \cdot \hat{\sigma}_i^{-}
\label{eq:jorwig_2}
\end{eqnarray}
where $\hat{S}_k$ are the sign change operators
acting locally on index $k$ and
their matrix representation reads
\begin{eqnarray}
\label{eq:mat_s_dof}
\bm{\sigma}^{+} =
\begin{pmatrix}
    0 & 0 \\
    1 & 0 \\
\end{pmatrix} ; \,\,\,\,\,\,\,
\bm{\sigma^{-}} =
\begin{pmatrix}
    0 & 1 \\
    0 & 0 \\
\end{pmatrix} ; \nonumber\\
\bm{S} =
\begin{pmatrix}
    1 & 0 \\
    0 & -1 \\
\end{pmatrix} ; \,\,\,\,\,\,\,
\bm{n} =
\begin{pmatrix}
    0 & 0 \\
    0 & 1 \\
\end{pmatrix} .
\end{eqnarray}
The equivalent form of the second quantized electronic Hamiltonian
\emph{in product form} is given by
\begin{align}
  \label{eq:ham_el_jw}
    \hat{H}_{e} = &
    \sum_{ij} h_{_{ij}}
        \left(
          \prod_{q=a+1}^{b-1} \hat{S}_q
        \right) \hat{\sigma}_i^{+} \hat{\sigma}_j^{-} \\\nonumber
     + &  \frac{1}{2}\sum_{ijkl} v_{_{ijkl}}
          \left( \prod_{q=a+1}^{b-1} \hat{S}_q
              \prod_{q^\prime=c+1}^{d-1} \hat{S}_{q^\prime}
          \right) \\\nonumber
       &   \mathrm{sgn}(j-i)\mathrm{sgn}(l-k)
    \hat{\sigma}_i^{+} \hat{\sigma}_j^{+} \hat{\sigma}_l^{-} \hat{\sigma}_k^{-}.
\end{align}
Here the indices $(a,b,c,d)$ correspond to the $(i,j,k,l)$ indices, but ordered from
smaller to larger and the $\mathrm{sgn}$ function is defined as
\begin{align}
    \mathrm{sgn}(x) =
        \begin{cases}
            1, & \text{if } x\geq 0\\
           -1, & \text{otherwise}.
\end{cases}
\end{align}

\subsubsection{Representations of the SQR wavefunction and operator}

There are two limiting wavefunction \emph{ansätze} when applying the
MCTDH-SQR formalism to fermionic systems:
(i) Each spin degree of freedom (S-DOF) is described by one SPF. The
\emph{second quantized}
wavefunction is then described by a single Hartree product.
(ii) Each S-DOF is described by two SPFs. In this case the wavefunction has
$2^M$ configurations, is exact, and spans all states of the
underlying Fock space.
Limiting case (i) is a poor representation of correlated states, whereas case
(ii) becomes quickly unaffordable as the number of spin-orbitals increases.
Therefore, the only practical alternative to apply MCTDH-SQR to fermions is to group
primitive spin degrees of freedom together, either through
mode combination or through the generalization to ML-MCTDH-SQR.
Grouping two or more S-DOFs the total Fock
space becomes
\begin{eqnarray}
 \mathcal{F}(M) = f_1(m_1) \otimes f_2(m_2) \otimes \dots \otimes f_d(m_d)
\end{eqnarray}
where $f_\kappa(m_\kappa)$ denotes the sub-Fock space of $m_\kappa$ S-DOFs with 2$^{m_\kappa}$ Fock states
and
\begin{eqnarray}
 M = \sum_{\kappa=1}^{d} m_\kappa.
\end{eqnarray}
In MCTDH language, $f_\kappa(m_\kappa)$ corresponds to the primitive space of the $\kappa$-th
logical coordinate.

One can go one step further and represent the states of the sub-Fock space
$f_\kappa(m_\kappa)$ as a new primitive degree of freedom.
We refer to this representation of the primitive degrees of freedom
as Fock space DOF (FS-DOF).
\begin{figure}[t]
\begin{center}
\includegraphics[width=5.5cm]{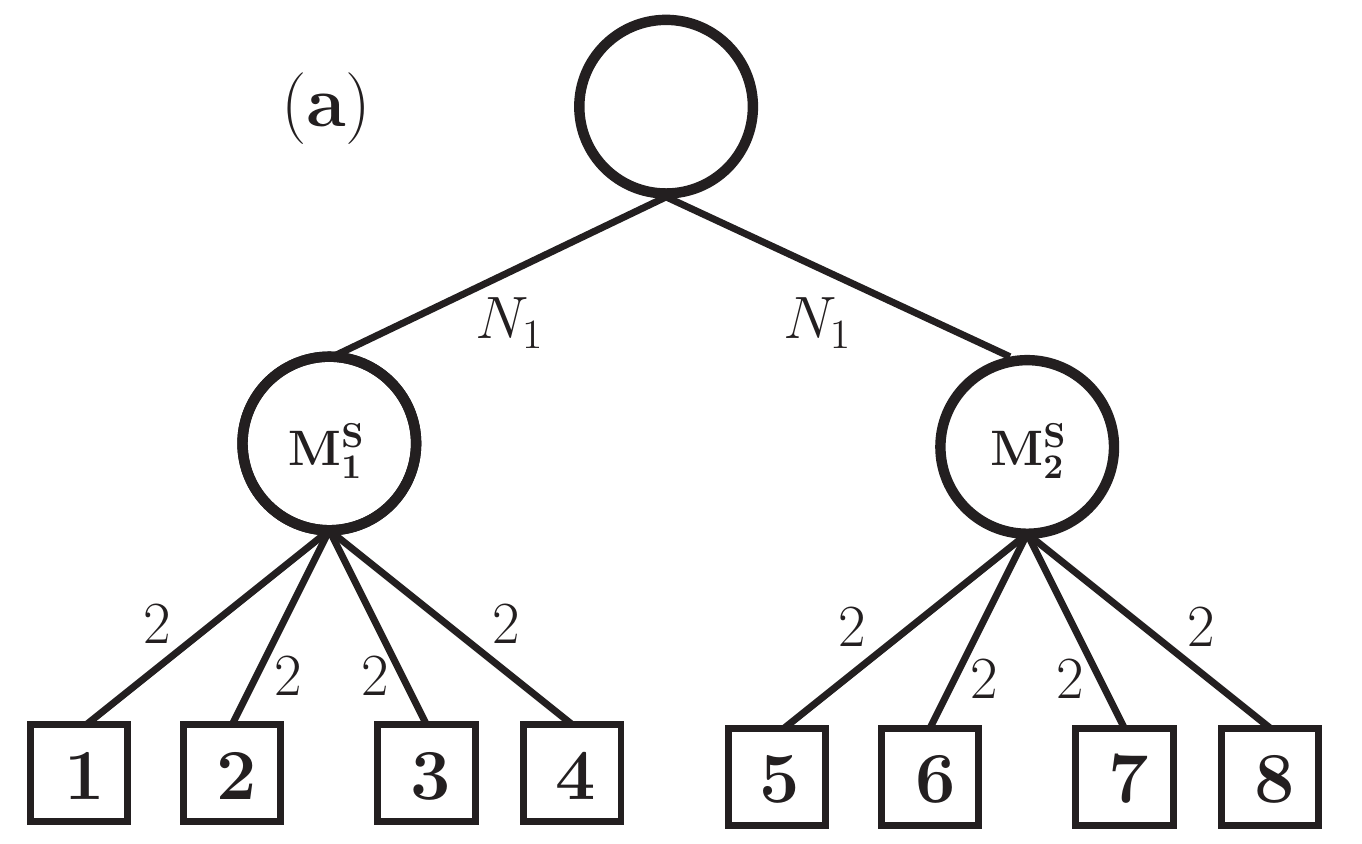}
\includegraphics[width=2.8cm]{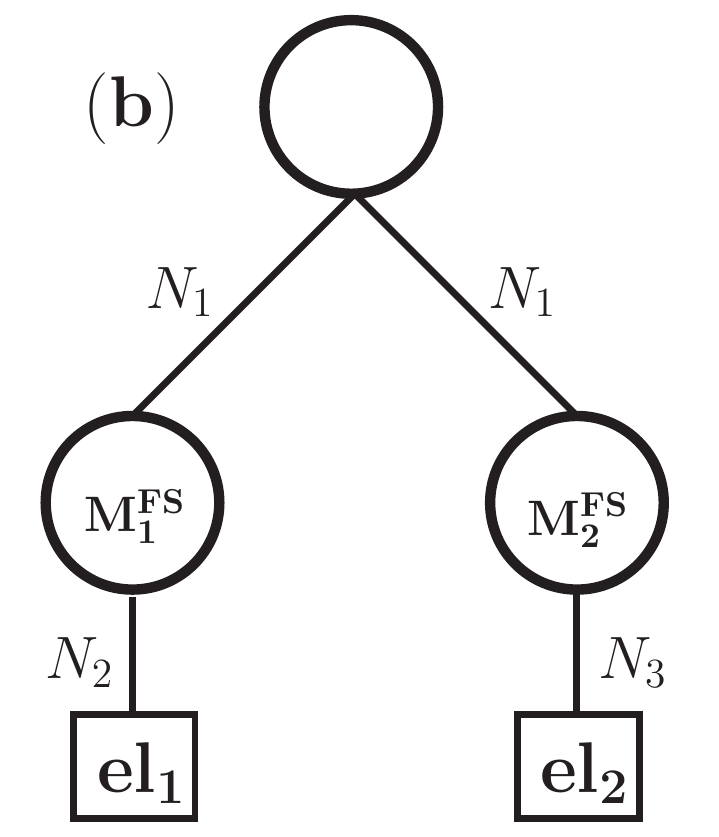}
\caption{Tree structures for the MCTDH electronic wavefunction
    containing 8 spin orbitals.
(a) MCTDH wavefunction tree, in which spin orbitals are considered as
    the primitive electronic DOFs (S-DOF).
(b) MCTDH wavefunction tree, in which the Fock
    space of the 4 combined spin orbitals is
    considered as a single electronic DOF (FS-DOF).
    }
\label{fig:s_fs_dof}
\end{center}
\end{figure}
%
There are several advantages when using FS-DOFs over S-DOFs:
\begin{enumerate}
\item The Fock space of each FS-DOFs can be statically pruned before the
calculation. This static pruning is not possible in the S-DOF representation.
The pruning of the Fock space is equivalent to removing unwanted grid points
from a multidimensional grid
in a discrete variable representation (DVR)~\cite{bac86:4594}.
A similar strategy, termed adiabatic contraction,
was used by Wang and Thoss in their applications using the
ML-MCTDH-SQR method~\cite{wan13:7431, wan16:164105}. The criteria to
prune the primitive space are different though. Whereas they employ an energetic cutoff,
we consider what are the possible electronic occupations of the subspace according to
various criteria.
As an example, let us consider a system containing 8 spin orbitals that are divided into two
groups, each with 4 spin orbitals. The MCTDH wavefunction trees for S-DOF
and FS-DOF are given in Fig. \ref{fig:s_fs_dof}.
For S-DOFs, each DOF has two primitive states $|0\rangle$ and $|1\rangle$
and the Fock space spanned by each combined mode
$M_1^{FS}$ and $M_2^{FS}$ mode, Fig. \ref{fig:s_fs_dof}(a), is
\begin{align}
    \mathcal{F}(4) \ni &
    (|0\rangle \oplus |1\rangle)_1\, \otimes
    (|0\rangle \oplus |1\rangle)_2\, \otimes\\\nonumber
    & (|0\rangle \oplus |1\rangle)_3\, \otimes
    (|0\rangle \oplus |1\rangle)_4
\end{align}
with $2^4 = 16$ Fock states.
In the FS-DOF case, each of the $M_1^{FS}$ and $M_2^{FS}$ degrees of freedom,
Fig. \ref{fig:s_fs_dof}(b), is spanned, in principle, by the 16 original Fock
states.
In practice, we one can prune the FS-DOF by either allowing only a certain range
of electronic occupations in each FS-DOF, by further removing certain
        occupation number states that are considered chemically irrelevant (for
        example based on energetic arguments), or
both.
For the above mentioned example, if the spin orbitals are energetically ordered
with respect to a pre-existing first quantization mean-field calculation,
then in the lowest energy configuration of the mean-field all the orbitals
in $el_1$ are occupied ($|1, 1, 1, 1\rangle$) and all orbitals
in $el_2$ are empty ($|0, 0, 0, 0\rangle$).
Now, if one decides to allow only upto two holes in $el_1$,
the occupation number states of the $\mathcal{H}(4, 0)$
and $\mathcal{H}(4, 1)$ Hilbert spaces are removed from the corresponding
$\mathcal{F}(4)$ Fock space, which shrinks from 16 to 9 occupation number
states.
Equivalently, one would in this case also limit the maximum number of electrons
in the $el_2$ Fock space to two, thus eliminating from it the occupation number
states of the $\mathcal{H}(4, 3)$ and $\mathcal{H}(4, 4)$ Hilbert spaces.
%


%
\item The operators acting on FS-DOFs are very sparse and correspond to pure
    mappings. This is probably the most important reason for introducing FS-DOFs
    and the one that can lead to the largest efficiency gains in computations.
    To illustrate this, let us consider again a simple example where the
    (2$\times$2) matrices
\begin{align}
\label{eq:ex_fs_dof1}
    \bm{\sigma^{+}}_{(1)} \bm{S}_{(2)} \bm{S}_{(3)} =
\begin{pmatrix}
    0 & 0 \\
    1 & 0 \\
\end{pmatrix}_{(1)}
\begin{pmatrix}
    1 & 0 \\
    0 & -1 \\
\end{pmatrix}_{(2)}
\begin{pmatrix}
    1 & 0 \\
    0 & -1 \\
\end{pmatrix}_{(3)}
\end{align}
    are applied to a sequence of three combined spin degrees of freedom. The
    (8$\times$8) matrix performing the same operation on the
    corresponding FS-DOFs reads
\begin{align}
\label{eq:ex_fs_dof2}
\begin{pmatrix}
    0 & 0 & 0 & 0 & 0 & 0 & 0 & 0 \\
    0 & 0 & 0 & 0 & 0 & 0 & 0 & 0 \\
    0 & 0 & 0 & 0 & 0 & 0 & 0 & 0 \\
    0 & 0 & 0 & 0 & 0 & 0 & 0 & 0 \\
    1 & 0 & 0 & 0 & 0 & 0 & 0 & 0 \\
    0 & -1 & 0 & 0 & 0 & 0 & 0 & 0 \\
    0 & 0 & -1 & 0 & 0 & 0 & 0 & 0 \\
    0 & 0 & 0 & 1 & 0 & 0 & 0 & 0 \\
\end{pmatrix}.
\end{align}
\emph{All} such combined matrix operators have the following properties:
there is at most one entry different than 0 in each row and column, which can be
either equal to $1$ or $-1$. In an $n\times n$ matrix, at most $n$ entries are
different than 0. These properties can be used to greatly increase the
efficiency of the multiplication of these matrices to vectors of coefficients
(representing an MCTDH single particle function). These operations can be
performed by dedicated subroutines by mapping coefficients and avoiding actual
number multiplications and summations. Note, for example, that the combination
of 10 spin-orbitals into one FS-DOF leading to a Fock space with $1024$
occupation number states results in matrix operators with at most $\approx$ 1/1000
non-zero entries, and even larger FS-DOFs are possible.

\item The pruning of the FS-DOF spaces can result in fewer product terms of the
    second quantized Hamiltonian~(\ref{eq:ham_el_jw}).
    This is the case for product terms,
    e.g. of the form of Eq.~(\ref{eq:ex_fs_dof1}),
    that in the original sub-Fock space link configurations that in the FS-DOF
    have been
    pruned.
\end{enumerate}


\subsection{Non-adiabatic quantum dynamics in second quantization}
\label{sec:theory:n-sqr}

One of the main purposes of this paper is to introduce and describe the
treatment of the coupled nuclear and electronic dynamics based on an SQR
representation of the electronic subsystem.
The full non-relativistic molecular Hamiltonian is given by
    \begin{align}
        \label{eq:hamTot1}
        \hat{H} = \hat{T}_n + \hat{W}_n + \hat{H}_e
    \end{align}
where the nuclear kinetic energy ($\hat{T}_n$) and the nuclear-nuclear potential energy
($\hat{W}_n $) operators are given by
    \begin{align}
        \label{eq:Tn}
        \hat{T}_n = \sum_{\alpha=1}^N -\frac{\hbar^2}{2m_\alpha} \vec{\nabla}_\alpha^2
    \end{align}

    and

    \begin{align}
        \label{eq:Wn}
        \hat{W}_n = \sum_{\alpha,\beta=1}^N
        \frac{Q_\alpha Q_\beta e^2}{|\vec{R}_\alpha-\vec{R}_\beta|},
    \end{align}
    respectively.
Here, the indices $\alpha,\beta$ run over all nuclear DOFs.
In this section it will be convenient to express the
electronic part ($\hat{H}_e$) of the total
molecular Hamiltonian in terms of second-quantized fermionic field operators
$\hat{\psi}^\dagger(\vec{x})$ and $\hat{\psi}(\vec{x})$
    \begin{align}
        \label{eq:He}
        \hat{H}_e & = \int\!\! dr_1^3\;
            \hat{\psi}^\dagger(\vec{x_1})
            \left(
                -\frac{\hbar^2}{2m_e}\vec{\nabla}^2
                + \sum_{\alpha=1}^{N}\frac{-Q_\alpha e^2}{|\vec{R}_\alpha-\vec{r}_1|}
            \right)
            \hat{\psi}(\vec{x_1}) \\\nonumber
            & +
            \int\!\! dr_1^3
            \int\!\! dr_2^3\;
            \hat{\psi}^\dagger(\vec{x_1})
            \hat{\psi}^\dagger(\vec{x_2})
            \frac{e^2}{|\vec{r}_1-\vec{r_2}|}
            \hat{\psi}(\vec{x_2})
            \hat{\psi}(\vec{x_1}), \\\nonumber
    \end{align}
    where $\vec{x}_j$
    denotes the spatial and spin coordinates of the fermionic particles.
    %
    %
    The simplest situation corresponds to the case in which
    the field operators are
    expanded in a basis of electronic spin-orbitals that are
    \emph{independent} of the nuclear positions
    \begin{align}
        \label{eq:fieldOpsA1}
          \hat{\psi}^\dagger(\vec{x}) & = \sum_{j=1}^F \chi_j^*(\vec{x})
          \hat{a}_j^\dagger\\
        \label{eq:fieldOpsA2}
          \hat{\psi}(\vec{x})         & = \sum_{j=1}^F \chi_j(\vec{x})
          \hat{a}_j.
    \end{align}
    Hamiltonian~(\ref{eq:hamTot1}-\ref{eq:He}) can be re-expressed using the
    mode expansions~(\ref{eq:fieldOpsA1}, \ref{eq:fieldOpsA2})
    resulting in
    \begin{align}
        \label{eq:hamTot}
        \hat{H} = \hat{T}_n + \hat{W}_n +
        \sum_{ij}^F h_{ij}(\mathbf{R}) \hat{a}_i^{\dagger} \hat{a}_j +
            \frac{1}{2}\sum_{ijkl}^F v_{ijkl}
            \hat{a}_i^{\dagger} \hat{a}_j^{\dagger}
            \hat{a}_l \hat{a}_k,
    \end{align}
    where
    \begin{align}
        \label{eq:hij1}
        h_{ij}(\mathbf{R}) =
           \langle\chi_i(\vec{x}_1)|
                \frac{-\vec{\nabla}^2}{2} +
                \sum_{\alpha=1}^{N}\frac{-Q_\alpha}{|\vec{R}_\alpha-\vec{r}_1|}
           |\chi_j(\vec{x}_1)\rangle
    \end{align}
    and
    \begin{align}
        \label{eq:vijkl1}
        v_{ijkl} =
           \langle\chi_i(\vec{x}_1)\chi_j(\vec{x}_2)|
                \frac{1}{|\vec{r}_1-\vec{r}_2|}
           |\chi_k(\vec{x}_1)\chi_l(\vec{x}_2)\rangle
    \end{align}
    are the one- and two-body integrals of the electronic subsystem,
    respectively, and the coupling of the electrons and nuclei occurs only through
    the one-body electronic term.

    In applications to molecular systems it may be more convenient to represent
    the electronic problem in a single particle basis that depends on the
    nuclear positions, thus removing to a large extent the artificial
    correlations among the positions of the nuclei and electrons in space and
    thus keeping the number of electronic spin-orbitals as small as possible.
    In such a basis, the same field operators~(\ref{eq:fieldOpsA1},
    \ref{eq:fieldOpsA2}) are expanded as
    \begin{align}
        \label{eq:fieldOpsB1}
          \hat{\psi}^\dagger(\vec{x}) & = \sum_{j=1}^F
          \varphi_j^*(\vec{x};\mathbf{R}) \hat{b}_j^\dagger(\mathbf{R})\\
        \label{eq:fieldOpsB2}
          \hat{\psi}(\vec{x}) & = \sum_{j=1}^F
          \varphi_j(\vec{x};\mathbf{R}) \hat{b}_j(\mathbf{R}).
    \end{align}
    At first sight it may look now as if one will obtain again
    Hamiltonian~(\ref{eq:hamTot}), but
    now expressed in the nuclear-dependent basis.
    This is not the case because the occupation number states
    carry an implicit nuclear-positon dependency (cf. Eq.~(\ref{eq:nstate_vac}))
    that
    results in non-adiababtic coupling (NAC) terms of kinetic nature:
    \begin{align}
        \label{eq:nac_T}
        \langle{\{\vec{m}\}}| \hat{T}_n |{\{\vec{n}\}}\rangle & =
        \hat{T}_n\delta_{\vec{m}\vec{n}} \\\nonumber
        & -\frac{1}{M}
        \langle{\{\vec{m}\}}| \boldsymbol{\nabla} |{\{\vec{n}\}}\rangle
        \boldsymbol{\nabla} \\\nonumber
        & -\frac{1}{2M}\langle{\{\vec{m}\}}| \boldsymbol{\nabla}^2
        |{\{\vec{n}\}}\rangle.
    \end{align}

    To evaluate the SQR NACs we first
    combine Eqs.~(\ref{eq:fieldOpsA1}, \ref{eq:fieldOpsA2}) and
    (\ref{eq:fieldOpsB1}, \ref{eq:fieldOpsB1})
    and express the
    the position-dependent creation (annihilation) operators,
    in terms of the
    position-independent operators as
    \begin{align}
        \label{eq:trafo1}
            \hat{b}_j(\mathbf{R}) & = \sum_{k=1}^F
            \langle \varphi_j(\mathbf{R}) | \chi_k \rangle \hat{a}_k
        \\
        \label{eq:trafo2}
            \hat{b}_j^\dagger(\mathbf{R}) & = \sum_{j=1}^F
            \langle \chi_k | \varphi_j(\mathbf{R}) \rangle \hat{a}^\dagger_k,
    \end{align}
    respectively,
    and then express the occupation number representation kets starting
    from the vacuum state as
    \begin{align}
        \label{eq:nstate_vac}
        |{\{\vec{n}\}}\rangle = \prod_{k=1}^F
        \left( \hat{b}_k^\dagger \right)^{n_k} |0\rangle.
    \end{align}
    Here and in the following we drop the $R$-dependence of the
    $\hat{b}_k$ and $\hat{b}_k^\dagger$ for notational simplicity. Note that
    $\hat{b}_k$ and $\hat{b}_k^\dagger$ act on the Fock space basis
    similarly as $\hat{a}_k$ and $\hat{a}_k^\dagger$, and in practice they are
    replaced by the spin matrices in Eqs.(\ref{eq:jorwig_2},\ref{eq:mat_s_dof}).
    Making use of the expressions above one arrives at
    \begin{widetext}
    \begin{align}
        \label{eq:nac1devel}
        \langle{\{\vec{m}\}}| \boldsymbol{\nabla} |{\{\vec{n}\}}\rangle & =
        \langle{\{\vec{m}\}}|
        \boldsymbol{\nabla}
        \prod_{l=1}^F
        \left( \hat{b}_l^\dagger \right)^{n_l}
        |0\rangle
        \\ \nonumber
        & =
        \langle{\{\vec{m}\}}|
        \sum_l n_l
        \left[
            \prod_{k<l}(\hat{b}_k^\dagger)^{n_k}
        \right]
        \left[
            \sum_q \langle\chi_q|\boldsymbol{\nabla}|\varphi_l(\mathbf{R})\rangle
            \hat{a}_q^\dagger
        \right]
        \left[
            \prod_{k>l}(\hat{b}_k^\dagger)^{n_k}
        \right]
        |0\rangle
        \\ \nonumber
        & =
        \langle{\{\vec{m}\}}|
        \sum_{rl} n_l
        \left[
            \prod_{k<l}(\hat{b}_k^\dagger)^{n_k}
        \right]
        \left[
            \sum_q
            \langle\chi_q|
            \varphi_r(\mathbf{R})\rangle \langle \varphi_r(\mathbf{R})|
            \boldsymbol{\nabla}|\varphi_l(\mathbf{R})\rangle
            \hat{a}_q^\dagger
        \right]
        \left[
            \prod_{k>l}(\hat{b}_k^\dagger)^{n_k}
        \right]
        |0\rangle
        \\ \nonumber
        & =
        \langle{\{\vec{m}\}}|
        \sum_{rl} n_l
        \left[
            \prod_{k<l}(\hat{b}_k^\dagger)^{n_k}
        \right]
        \left[
            \langle \varphi_r(\mathbf{R})|
            \boldsymbol{\nabla}|\varphi_l(\mathbf{R})\rangle
            \hat{b}_r^\dagger
        \right]
        \left[
            \prod_{k>l}(\hat{b}_k^\dagger)^{n_k}
        \right]
        |0\rangle,
    \end{align}
    \end{widetext}
    where the nuclear derivative operator acts now on the basis of position
    dependent spin-orbitals.
    Finally, the multiplicative $n_l$ term in the last line of
    Eq.~(\ref{eq:nac1devel}) can be substituted by $b_l(b_l^\dagger)^{n_l}$,
    leaving the matrix element unaltered and allowing to re-introduce
    the ket $|{\{\vec{n}\}}\rangle$ on the right side
    \begin{align}
        \label{eq:nac1devel_final}
        \langle{\{\vec{m}\}}| \boldsymbol{\nabla} |{\{\vec{n}\}}\rangle & =
        \sum_{rl}
        \langle \varphi_r(\mathbf{R})|\boldsymbol{\nabla}|\varphi_l(\mathbf{R})\rangle
        \langle{\{\vec{m}\}}|
            \hat{b}_r^\dagger
            \hat{b}_l
        |{\{\vec{n}\}}\rangle.
    \end{align}
    The second order non-adiabatic coupling term is obtained in the same manner
    and reads
    \begin{align}
        \label{eq:nac2devel_final}
        \langle{\{\vec{m}\}}| \boldsymbol{\nabla}^2 |{\{\vec{n}\}}\rangle & =
        \sum_{rl}
        \langle \varphi_r(\mathbf{R})|\boldsymbol{\nabla}^2|\varphi_l(\mathbf{R})\rangle
        \langle{\{\vec{m}\}}|
            \hat{b}_r^\dagger
            \hat{b}_l
        |{\{\vec{n}\}}\rangle.
    \end{align}

    Using the matrix elements in
    Eqs.~(\ref{eq:nac1devel_final},\ref{eq:nac2devel_final}),
    the total molecular Hamiltonian~(\ref{eq:hamTot1})
    takes the form
    \begin{widetext}
    \begin{align}
        \label{eq:ham_nuc_sqr}
        \hat{H} =
        \hat{T}_n
        +
        \hat{W}_n
        +
        \sum_{pq}^F
            \left[
                {h}_{pq} (\mathbf{R})
                -\frac{1}{M}\mathbf{D}_{pq}(\mathbf{R})\cdot\boldsymbol{\nabla}
                -\frac{1}{2 M} G_{pq}(\mathbf{R})
            \right]
        \hat{b}_p^{\dagger} \hat{b}_q
        +
        \frac{1}{2}\sum_{pqrs}^F {v}_{pqrs}(\mathbf{R})
        \hat{b}_p^{\dagger} \hat{b}_q^\dagger
        \hat{b}_s \hat{b}_r
    \end{align}
    \end{widetext}
    in the basis of position-dependent spin-orbitals,
    where
    \begin{align}
        \label{eq:nac1_expression}
        \mathbf{D}_{pq} =
        \langle \varphi_p(\mathbf{R})|\boldsymbol{\nabla}|\varphi_q(\mathbf{R})\rangle
    \end{align}
    and
    \begin{align}
        \label{eq:nac2_expression}
        G_{pq} =
        \langle \varphi_p(\mathbf{R})|\boldsymbol{\nabla}^2|\varphi_q(\mathbf{R})\rangle
    \end{align}
    represent the first and second order non-adiabatic couplings that arise from
    the nuclear position dependence of the electronic orbital basis, and
    the integrals ${h}_{pq}$ and ${v}_{pqrs}$ are evaluated in the
    $|\varphi_q(\mathbf{R})\rangle$ basis.

    %
    %
    All degrees of freedom in the molecular Hamiltonian~(\ref{eq:ham_nuc_sqr}),
    nuclei positions and electronic occupations, are distinguishable.
    We refer to the combination of a first quantized representation for the
    nuclei and a second quantized representation of the fermionic
    subsystem, Hamiltonian~(\ref{eq:ham_nuc_sqr}), as the N-SQR framework.
    N-SQR enables the application of various forms of tensor decompositions to
    the corresponding wavefunction, such as hierarchical tensor contractions and
    matrix product-state representations.
    Hence, MCTDH and its ML-MCTDH generalization, which can be
    regarded as a hierarchical Tucker decompositions of the wavefunction, can be
    readily applied to Hamiltonian Eq.~(\ref{eq:ham_nuc_sqr}) to describe the
    mixed non-adiabatic dynamics of nuclei and electrons without, in principle,
    further modification.
    Nonetheless, practical aspects need to be considered in N-SQR calculations,
    which are the subject of the next section.

    For completeness, Appendix~\ref{ap:bh} describes formally the application of
    the Hamiltonian~(\ref{eq:ham_nuc_sqr}) to a Born-Huang-like expansion of the
    wavefunction, although this alternative is not used in the actual
    calculations reported here.

\subsection{Practical aspects of N-SQR and implementation within MCTDH}
\label{sec:theory:impl}

\subsubsection{Cutoff strategies for Hamiltonian terms}


    A major obstacle in the \emph{ab initio} description of molecules
    within the SQR framework is the large number of
    Hamiltonian terms to be considered, mostly due to the
    four-index Coulomb integrals $v_{ijkl}$.
    In this work, we apply two strategies to mitigate this problem:

    -- Frozen core approximation: --
    This strategy is widely used in static multi-configuration
    electronic structure calculations.
    Low lying molecular orbitals, e.g. linear
    combinations of atomic $1s$ orbitals, are energetically
    well separated from the valence space and, for most practical purposes,
    remain occupied at all times.
    Removing them leads to a reduction of the number of degrees of freedom in
    a similar manner as when introducing
    reduced dimensionality models in nuclear dynamics, and also to a
    reduction in the number of Hamiltonian terms.
    Restricting the lowest energy
    spin-orbitals to always being occupied and dividing the orbital space
    into two groups, core orbitals
    ($i,j, \dots$) and active orbitals ($a,b,\dots$),
    the energy of the core region is given by
    \begin{eqnarray}
        \label{eq:core2}
        E^{core}(\mathbf{R}) & =
        \sum_{i} h_{ii}(\mathbf{R})
        + \frac{1}{2} \sum_{ij} 
        \left( v_{ijij}(\mathbf{R}) - v_{ijji}(\mathbf{R})  \right),
    \end{eqnarray}
    which contributes to the Hamiltonian as a potential energy surface for the
    nuclei.
    The electrons in the active orbitals $(a,b)$ still interact with the filled
    core and the one-body transfer integrals $h_{ab}(\mathbf{R})$
    need to be adjusted
    accordingly as
    \begin{align}
        \label{eq:one_mod}
        & \tilde{h}_{ab}(\mathbf{R}) = 
        h_{ab}(\mathbf{R}) \\\nonumber
        & + \frac{1}{2} \sum_i
        \left( v_{iaib}(\mathbf{R}) +
               v_{aibi}(\mathbf{R}) -
               v_{iabi}(\mathbf{R}) -
               v_{aiib}(\mathbf{R}) \right).
    \end{align}
    A similar strategy can be applied if a group of spin-orbitals can be assumed
    to be, at most, singly occupied at all times. This would be the case, e.g.,
    in single ionization processes, where, per definition, only one electron
    reaches the orbitals that describe the outgoing particle. In this group of
    orbitals, only one-body terms of the Hamiltonian need to be retained.
    Spin-orbitals that are considered to never become populated can be removed
    from the calculation without further consideration.

    -- Selective cutoff value for Hamiltonian terms: -- It is common in MCTDH
    calculations to set a cutoff and remove small Hamiltonian terms, e.g., of
    order $10^{-7}$ or smaller before time propagations are performed. We have
    observed that this strategy is not so convenient with the electronic SQR
    Hamiltonian because it consists of thousands of very small contributions
    which, when removed, lead to noticeable effects.
    For SQR we proceed in a slightly modified way. We apply a predefined cutoff
    only to 2-body terms of 3-index and 4-index type, e.g. $v_{1,2,3,3}$ or
    $v_{1,2,3,4}$.
    The rationale behind this strategy is that only one- and two-body
    terms with up to two different indices, the Coulomb and exchange integrals,
    contribute to the energy of an occupation number state
    $\langle \{\vec{n}\}| \hat{H}_e | \{\vec{n}\}\rangle$, so we keep them all.
    Instead, 3-index and 4-index integrals participate only in the off-diagonal
    terms that connect different occupation number states.
    We have observed that this strategy to be effective in reducing the size of the
    Hamiltonian in a more or less controllable way. This will be illustrated
    later in the discussion of the results.


\subsubsection{Diabatization of orbitals}

    One important aspect of static multi-configuration electronic structure
    calculations aimed at the generation of potentials and NAC coupling elements
    is the introduction of a basis of molecular orbitals that is \emph{as
    diabatic as possible}~\cite{pac91:6668,%
    dom93:362,%
    gar97:161,%
    nak01:10353,%
    nak02:5576,%
    yan13:84}.
    In this case, the NAC couplings between adiabatic electronic states are
    captured, practically in their entirety, by the configuration interaction
    (CI) expansion vectors~\cite{pac91:6668}.
    In the nuclear-SQR framework, the advantage of using (quasi-)diabatic
    orbitals is similar. The differential terms in
    Hamiltonian~(\ref{eq:ham_nuc_sqr}) can be neglected and all the
    non-adiabatic effects are included in the time-evolution of the overall
    wavefunction.

    There are several diabatization procedures described in literature that produce
    (quasi-)diabatic molecular orbitals
    \cite{pac91:6668, dom93:362, gar97:161, nak01:10353, nak02:5576,yan13:84},
    for example imposing conformational uniformity at close-lying
    geometries~\cite{pac91:6668, dom93:362,nak01:10353, nak02:5576}
    or extracting the orbitals from
    state-averaged self-consistent field calculations~\cite{gar97:161}.
    In this work, we simply reorder the
    Hartree-Fock (HF) molecular orbitals obtained at each nuclear geometry to
    ensure their continuity.
    In our selected example this is possible because of two reasons. First, the
    coupled muclear-electronic dynamics involve a group of highly excited
    electronic configurations, whereas the electronic ground state is, to a very
    good approximation, mono-configurational along the whole set of internuclear
    distances.  Hence, the HF mean-field presents no discontinuities. Second, we
    are considering only one nuclear coordinate.
    In general, though, obtaining a good set of quasi-diabatic orbitals will be
    an important aspect of N-SQR calculations, which will need a more
    detailed consideration in future works.

    We need to rearrange the MOs such that each MO retains its character over
    all geometries.
    In short, we proceed as follows.
    We construct an overlap matrix between the MOs of two successive geometries
    $R, R_0$
    \begin{align}
        \label{eq:s_mo}
        \mathbf{S}_{[R,R_0]}^{MO}
        & = \bm{\Psi}_{[R]}^T \bm{\Psi}_{[R_0]} \\\nonumber
        & = \bm{C}_{[R]}^T \bm{\Phi}_{[R]}^T \bm{\Phi}_{[R_0]} \bm{C}_{[R_0]},
    \end{align}
    where the vector notation, e.g. for column vector of MOs
    $\bm{\Psi}_{[R]}^T
    = (|\varphi_1\rangle, |\varphi_2\rangle,\dots)_{[R]}^T$, is used,
    the subscript denotes the corresponding geometry,
    and the MOs in terms of atomic orbitals (AOs) are given as
    \begin{eqnarray}
        \bm{\Psi}_{[R]} = \bm{\Phi}_{[R]} \bm{C}_{[R]},
    \end{eqnarray}
    where $\bm{C}_{[R]}$ is the AO-MO transformation matrix with the molecular
    orbital coefficients in its columns.
    Since atomic orbitals are, in general, not orthogonal and we are interested
    in their overlap at different, but close, geometries, a direct
    multiplication of the coefficient matrices $\bm{C}_{[R]}$ in
    Eq.~(\ref{eq:s_mo}) assuming that they flank a unit matrix is not possible.
    Therefore, we introduce symmetrically
    orthogonalized atomic orbitals (SO)
    \begin{eqnarray}
        \bm{\chi}_{[R]} = \bm{\Phi}_{[R]} \bm{s}_{[R]}^{-\frac{1}{2}},
    \label{eq:so}
    \end{eqnarray}
    where $\bm{s}_{[R]}$ is the overlap matrix between AOs at geometry $R$.
    \begin{eqnarray}
        \bm{s}_{[R]} = \bm{\Phi}_{[R]}^T \bm{\Phi}_{[R]}
        \label{eq:overlap_ao}
    \end{eqnarray}
    Now, using Eq.~(\ref{eq:so}), we can rewrite the Eq.~(\ref{eq:s_mo}) as
    \begin{align}
        \mathbf{S}_{[R,R_0]}^{MO} & =
        \bm{C}_{[R]}^T \bm{s}_{[R]}^{\frac{1}{2}}
        \left( \bm{\chi}_{[R]}^T \bm{\chi}_{[R_0]} \right)
        \bm{s}_{[R_0]}^{\frac{1}{2}} \bm{C}_{[R_0]} \nonumber\\
        & \approx \bm{C}_{[R]}^T \bm{s}_{[R]}^{\frac{1}{2}}
                  \bm{s}_{[R_0]}^{\frac{1}{2}} \bm{C}_{[R_0]}
    \end{align}
    where we have approximated the overlap between the SOs of two successive
    geometries $\left( \bm{\chi}_{[R]}^T \bm{\chi}_{[R_0]} \right)\approx
    \bm{1}$ as a unit matrix.
    This is a good approximation as long as the $R$ and $R_0$ are close enough.
    The simple diabatization procedure consists now in rearranging the MOs at
    geometry $R$ (columns of $\bm{C}_{[R]}$) such that the largest matrix
    elements of the overlap matrix $\mathbf{S}_{[R,R_0]}^{MO}$ are found in the
    main diagonal, and in multiply a phase of $-1$ to the corresponding column
    of $\bm{C}_{[R]}$, if needed, to render the diagonal matrix elements
    positive.

\subsubsection{Generation of initial SPFs}

    An important aspect in MCTDH calculations is the generation of the initial
    SPFs in Eq.~(\ref{eq:spfs}).
    Recently, Weike and Manthe~\cite{wei20:034101} showed that if the initial
    SPFs are eigenfunctions of the local particle-number operator of the
    corresponding $k$-th sub-Fock space
    \mbox{$\hat{n}^{(k)} = \sum_{i_k}^{N_k} \hat{a}_{i_k}^\dagger
    \hat{a}_{i_k}$},
    the MCTDH EOM conserve the particle number of each SPF during the time
    propagation.
    This means, the initially occupied SPF(s) in each sub-Fock space (combined
    mode or FS-DOF) must correspond to an integer particular particle number.
    Moreover, the initially unoccupied SPF(s) must also each correspond to an
    integer particle number and their set has to span various particle numbers
    such that the mode is flexible enough to represent the evolution of the
    system in the whole Fock space.
    Therefore, the choice of both the initially occupied and unoccupied initial SPFs
    is very important in MCTDH-SQR calculations.
    At the moment, we select the initial SPFs
    \begin{align}
        \label{eq:iniSPFs}
        |\phi_{j_k}^{(k)} \rangle = \sum_{i_k}^{N_k} c_{i_k,j_k}^{(k)}
        |\{\vec{n}\}_{i_k}\rangle
    \end{align}
    by only mixing primitive occupation-number
    states that are eigenfunctions of the local particle number operator
    \mbox{$\hat{n}^{(k)} |\{\vec{n}\}_{i_k}\rangle = n_{i_k}^{(k)}
    |\{\vec{n}\}_{i_k}\rangle$}
    with the same number of particles $n_{i_k}^{(k)}$.

    In molecular problems, the choice of what particle-numbers are
    spanned by the $k$-th sub-Fock space can be made similarly as when
    considering restricted active spaces in electronic multi-configuration calculations.
    For example, one may consider an initial electronic configuration with a
    set of low energy occupied, and a set of empty (virtual) spin-orbitals, and
    one may assume that at most $m$ electrons have, at any time, been
    transferred between the initially occupied and unoccupied spaces.
    This type of strategy has the added advantage of limiting the primitive
    particle-number states when using FS-DOFs, in contrast to
    S-DOFs, where the primitive space is not affected by such considerations.

    Since, as pointed out by Manthe~\cite{wei20:034101}, the SPFs conserve the local
    particle number, and the propagation of the expansion coefficients of the
    $A$-vector conserves the total particle number, MCTDH-SQR calculations based
    on SPFs with an integer particle number have entries of the $A$-vector that
    stay equal to zero at all times.
    These coefficients refer to configurations of the whole Fock space
    perpendicular to the initially populated Hilbert space.
    Currently, our code does not take advantage of this fact, but specialized
    MCTDH-SQR implementations might take advantage of these considerations.
\section{Computational details}           \label{sec:computation}

    We apply the N-SQR framework in combination with MCTDH-SQR
    to the photofragmentation of the
    HeH$^{+}$ molecule upon excitation by XUV photons of about 30~eV.
    Following the photoexcitation process,
    it is known that non-adiabatic effects couple different fragmentation
    channels and result in quantum interferences that shape the
    channel-selected photodissociation cross section~\cite{dum09:165101}.
    The N-SQR calculations are compared with traditional first quantization
    Born-Huang~\cite{Born-Crystal} (FQBH) calculations based on the GBOA.
    We perform the FQBH calculations in the \emph{adiabatic} representation and
    thus we use the computed PES and NACs from \emph{ab initio} quantum
    chemistry directly and without further transformation to a quasi-diabatic
    electronic basis.
    Technical details on the FQBH quantum dynamics calculations in the adiabatic
    representation are provided in Appendix~\ref{ap:fq}.

    The uncontracted aug-cc-pVTZ~\cite{dun89:1007} atomic basis is used for both H and He to
    generate the underlying spin-orbitals for the N-SQR calculation.
    As already mentioned, we use the Hartree-Fock molecular orbitals (MO)
    as the one-particle basis used to construct the matrix
    elements in Hamiltonian.~(\ref{eq:ham_nuc_sqr}).
    The primitive orbital space consists of 10
    spin-orbitals (5 lowest-energy spatial MOs of $\Sigma$ symmetry
    in the $C_\infty$ point group)
    of which two are occupied in the HF reference configuration.
    The
    one-electron and two-electron matrix elements in the atomic orbital (AO) basis
    and the AO-MO transformation coefficients $\bm{C}_{[R]}$  are obtained using
    the PSI4 program package
    \cite{par17:3185}.
    The columns of the $\bm{C}_{[R]}$ matrix are then sorted out for all
    geometries using the orbital
    diabatization procedure outlined above and subsequently
    the AO to MO transformation of the integrals is performed
    using the standard expressions
     \begin{align}
        \langle i|\hat{h}|j \rangle_{[R]} = &
            \sum_{\mu\nu}C^{*}_{\mu i,[R]} C_{\nu j,[R]}
            \langle \mu|\hat{h}|\nu \rangle_{[R]} \\\nonumber
        \langle ij|\hat{v}|kl \rangle_{[R]} = &
            \sum_{\mu\nu\lambda\sigma}C^{*}_{\mu i,[R]} C^{*}_{\nu j,[R]} C_{\lambda k,[R]} C_{\sigma l,[R]}
            \langle \mu\nu|\hat{v}|\lambda\sigma \rangle_{[R]}
        \end{align}
    where $\mu$, $\nu$, $\lambda$, $\sigma$ are the AO and $i$, $j$, $k$, $l$
    are the MO indices.
    %
    In the N-SQR calculations we neglect the NAC terms related to the orbital
    basis, Eqs.~(\ref{eq:nac1_expression}) and (\ref{eq:nac2_expression}),
    and show later when comparing to non-adiabatic first
    quantization calculations including the full \emph{ab initio} couplings that
    this approximation is an excellent one in the present case.
    \begin{figure}[t!]
    \begin{center}
    \includegraphics[width=8.5cm]{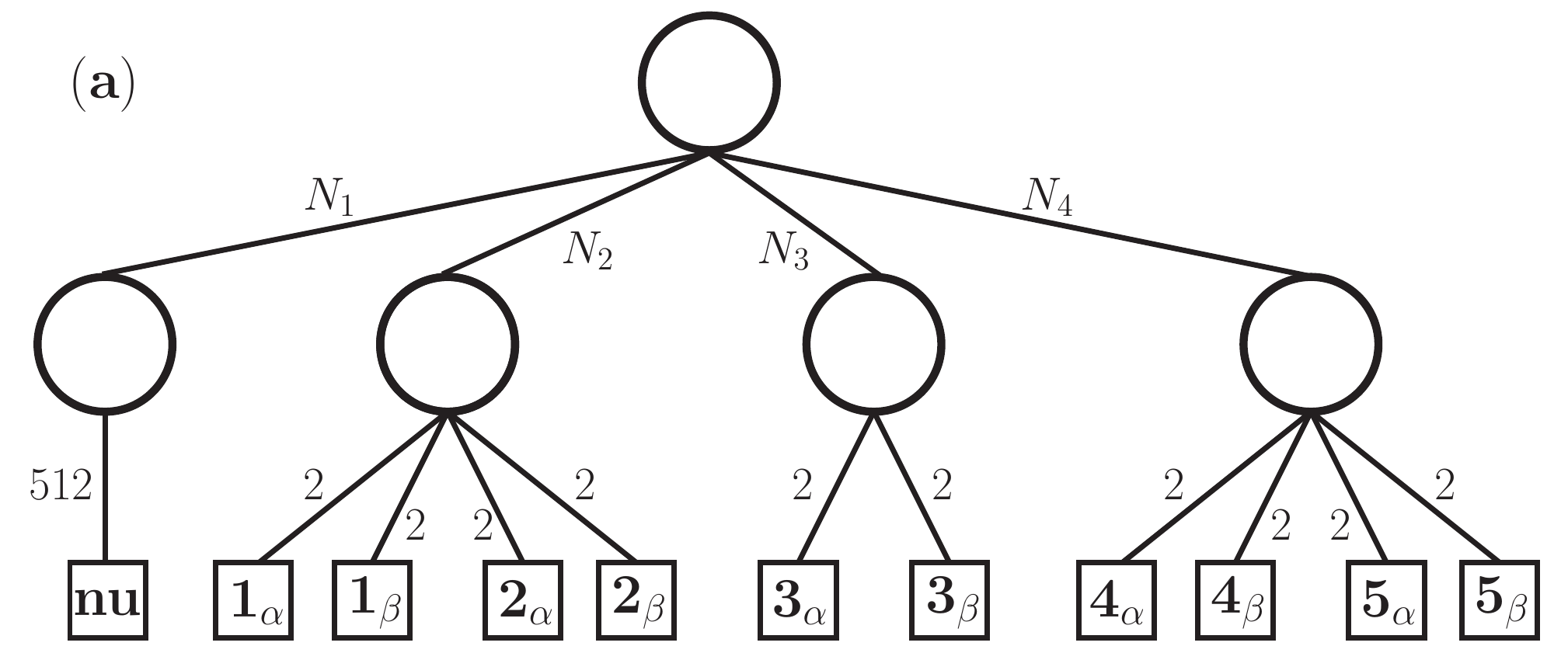}
    \hspace{2em}
    \includegraphics[scale=0.1, height=4cm]{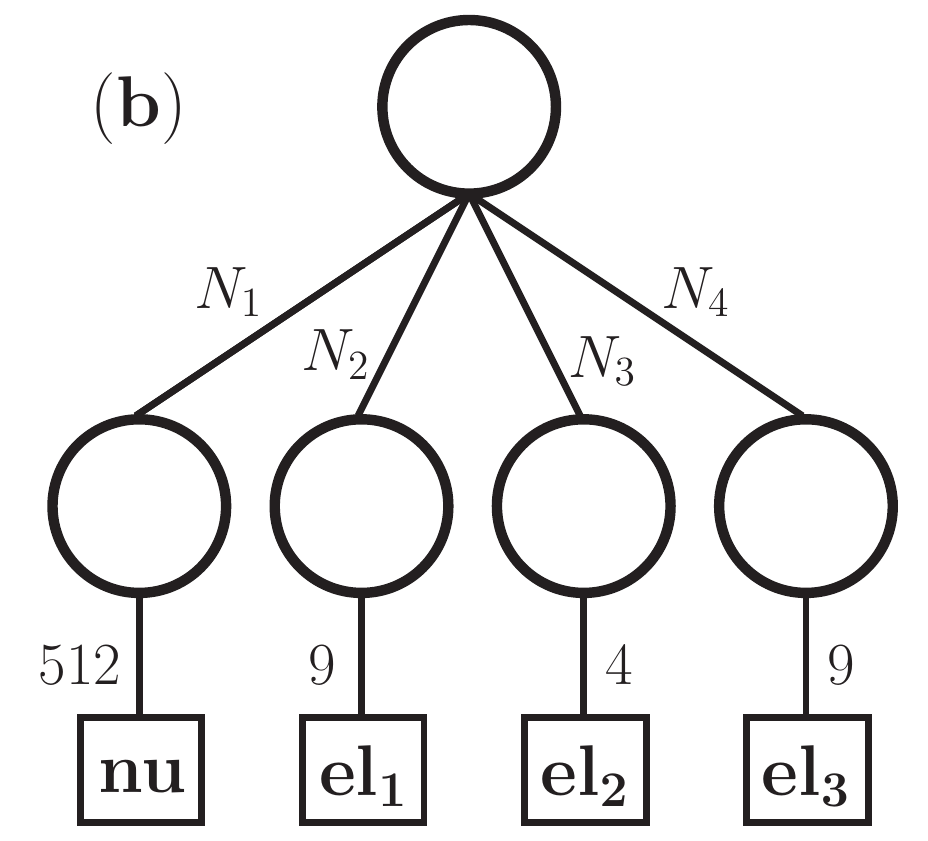}
    \vspace{4em}
    \includegraphics[scale=0.1, height=4cm]{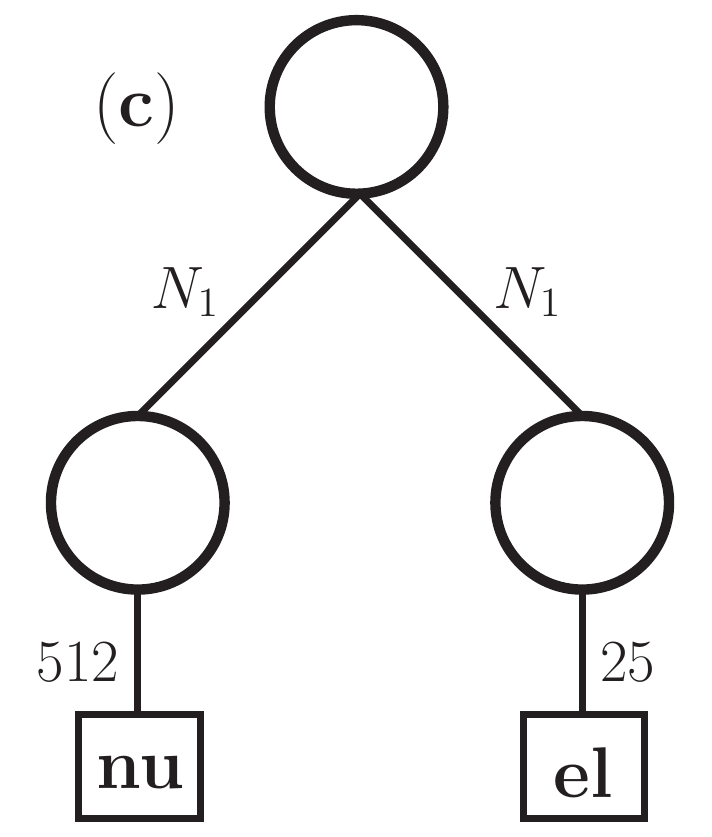}
    \caption{Tree structures for the MCTDH wavefunctions of the $HeH^{+}$. 
    (a) MCTDH wavefunction tree, in which spin orbitals are considered as
        the primitive electronic DOFs (S-DOF).
    (b) MCTDH wavefunction tree, in which the Fock space
        (consists of (4, 2, 4) spin orbitals) is considered as the
        electronic DOF (FS-DOF(I))
    (c) MCTDH wavefunction tree, in which the Fock space
        (consists of 10 spin orbitals) is considered as the
        electronic DOF (FS-DOF(II)). ``nu'' refers to the nuclear DOF and ``el$_i$'' to
        the FS-DOF.}
    \label{fig:s_fs_dof_heh+}
    \end{center}
    \end{figure}
    
    Three different representations
    of the electronic primitive basis of HeH$^{+}$
    are considered and
    compared. 
    These are termed
    S-DOF, FS-DOF(I) and FS-DOF(II), and their tree representations are
    illustrated in
    Fig.  \ref{fig:s_fs_dof_heh+}.
    On the one hand we use a S-DOF representation with three electronic modes
    formed by combining the alpha and beta spin orbitals of (1, 2), (3) and (4,
    5) spatial MOs (see Fig.~\ref{fig:s_fs_dof_heh+}a).
    The other extreme is FS-DOF(II), where all electronic degrees of freedom are
    combined into a single FS-DOF.
    In the latter case, due to the presence of two electrons of opposite spin in
    the system, the underlying Fock space with $2^{10}$ configurations shrinks
    to an FS-DOF with 25 particle-number states and the electronic operators in
    this space are represented hence by $25\times 25$ matrices.
    FS-DOF(I) is an intermediate case in which the S-DOFs in
    Fig.~\ref{fig:s_fs_dof_heh+}a are collected into FS-DOFs, which become the
    new electronic primitive degrees of freedom of the MCTDH wavefunction.

    The construction of the N-SQR input files for MCTDH, including calls to
    external electronic structure tools~\cite{par17:3185}, and the automatic
    generation of the corresponding input and operator files, is performed by
    dedicated modules written in Python, which will be made available in a future
    release of the  Heidelberg MCTDH program package
    \cite{mctdh:MLpackage}.
    The wavepacket propagation in both the traditional and N-SQR calculations
    is done as well using the Heidelberg MCTDH program suite
    \cite{mctdh:MLpackage}.

\section{Results and Discussion}           \label{sec:results}

\subsection{Electronic eigenenergies from SQR calculations}

\begin{figure}[t]
    \centering
    \begin{center}
    \includegraphics[width=8.5cm]{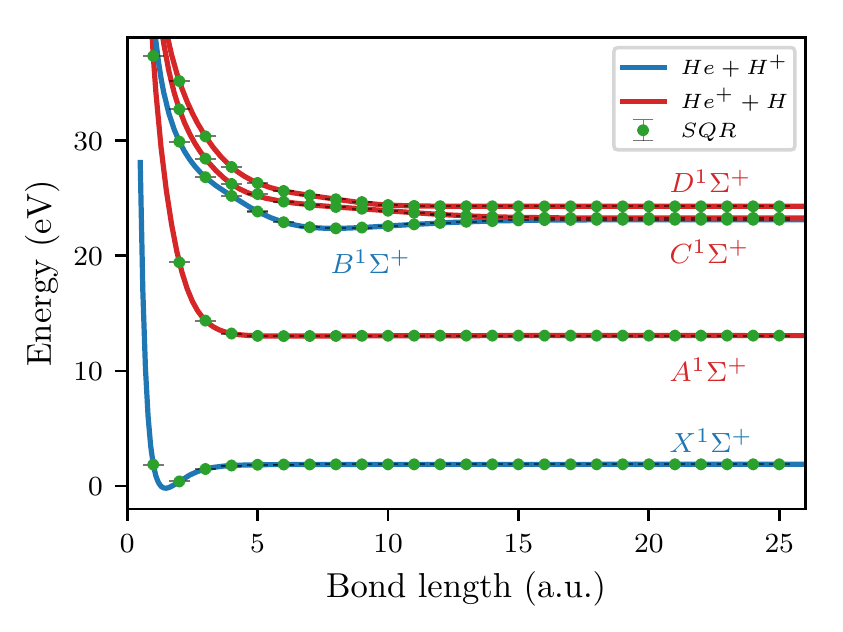}
    \caption{Comparison of first quantization full configuration interaction and
        second quantization representation potential energy surfaces.}
    \label{fig:comp_pes}
    \end{center}
    \end{figure}

    First of all, we apply the \emph{ab initio} MCTDH-SQR machinery
    to retrieve the
    correct adiabatic electronic energies, the potential energy curves,
    from the electronic SQR
    Hamiltonian at \emph{fixed nuclear geometries}.
    In the first quantization calculations, the adiabatic electronic energies
    for the lowest five $^1\Sigma^{+}$ states of HeH$^{+}$ are obtained
    through diagonalization of the full configuration interaction (FCI)
    Hamiltonian in the space of the 10 lowest energy HF
    spin-orbitals of $\Sigma$ symmetry, and are shown as solid lines
    in Fig.~\ref{fig:comp_pes}. The calculations of the potential energy
    curves were performed with the MOLPRO package~\cite{molpro}.
    In SQR, the PECs are obtained through time propagation of the
    MCTDH-SQR wavefunction
\begin{align}
 \label{init_elwf}
   |\Psi(t=0)\rangle
    &= | 1_\alpha, 1_\beta  \rangle \\\nonumber
    &+ | 1_\alpha, 2_\beta  \rangle
     - | 1_\beta,  2_\alpha \rangle \\\nonumber
    &+ | 1_\alpha, 3_\beta, \rangle
     - | 1_\beta,  3_\alpha \rangle \\\nonumber
    &+ | 1_\alpha, 4_\beta  \rangle
     - | 1_\beta,  4_\alpha \rangle \\\nonumber
    &+ | 1_\alpha, 5_\beta  \rangle
     - | 1_\beta,  5_\alpha \rangle.
\end{align}
    in the same spin-orbital space.
    The electronic wavefunction is represented with either one
    or three FS-DOFs (electronic part of the
    Figs.~\ref{fig:s_fs_dof_heh+}b and c, respectively)
    and the performed MCTDH-SQR calculations are numerically exact.
    The wavefunction~(\ref{init_elwf}) is spin-singlet and overlaps with all
    five $^1\Sigma^{+}$ states under consideration.
    The electronic eigenenergies are obtained from the maxima of the peaks in
    the power spectrum obtained from the Fourier transform of the
    autocorrelation function
    \begin{eqnarray}
        \sigma(E) = \frac{1}{\pi}Re\int_{0}^{\infty} e^{iEt}
        \langle \Psi|\Psi(t) \rangle dt.
    \end{eqnarray}
    The SQR electronic eigenenergies are shown as green marks
    superimposed in Fig.~\ref{fig:comp_pes}.
    The purpose of this comparison is to illustrate the correctness of our
    implementation and to hint at the possibility of computing \emph{ab
    initio} electronic energies and spectra in an SQR setting through time
    propagation.
    For example, an \emph{ab initio} time-dependent SQR propagation may serve
    to obtain ionization electronic spectra, and to calculate the
    electronic dynamics of molecular systems with a strong multi-configurational
    character, i.e. static correlation. Applications in these areas
    remain to be explored.

\subsection{Non-adiabatic dynamics in the N-SQR framework}


\subsubsection{Comparison between FQBH and N-SQR}

    Upon photoexcitation, the HeH$^{+}$ molecule can dissociate into two main
    fragmentation channels: He+H$^{+}$ and He$^{+}$+H.
    The five lowest adiabatic $^1\Sigma^{+}$ potential energy curves are shown
    in Fig.~\ref{fig:comp_pes} for comparison, where blue and red colors indicate
    the He+H$^{+}$ and He$^{+}$+H channels, respectively.

The photodissociation cross section of HeH$^{+}$ is obtained both in
the FQBH and N-SQR approaches using a complex absorbing potential (CAP)
of the form \mbox{$W(R)=-i\eta(R-R_c)^2$}
(for $R>R_c$) with \mbox{$R_c=20\,a_0$},
\mbox{$\eta=1.935\cdot 10^{-3}$}, and perfoming a flux
analysis of the propagated wavepacket~\cite{jae96:6778}.
For the comparison in this work we consider only the case that the light is polarized along
the molecular axis.
The propagated wavepacket is generated by application of the $\hat{M}^{(z)}$
dipole operator to the vibrational and electronic ground state of the system.
In the FQBH setting this operator reads
\begin{align}
    \label{eq:dipFQ}
    \hat{M}^{(z)}(R)=\sum_s \mu_{0s}^{(z)}(R)\Big(|0\rangle\langle s| + |s\rangle\langle
    0|\Big),
\end{align}
where $s$ refers to excited adiabatic electronic states and $\mu_{0s}(R)$ is the
corresponding transition-dipole matrix element.
In the N-SQR setting the dipole operator reads
\begin{align}
    \label{eq:dipNSQR}
    \hat{M}^{(z)}(R) = \sum_{ij}
    \langle \varphi_i(R) | \hat{\mu} | \varphi_j(R) \rangle
    \hat{b}_{i}^{\dagger}\hat{b}_j,
\end{align}
which is a one-body operator and the MO dipole integrals are obtained using the
integral engine of the PSI4 electronic structure package \cite{par17:3185}.


In the FQBH calculation, the fragmentation channel-resolved cross sections can
be obtained by projecting, before the flux analysis, the propagated wavefunction
onto each electronic state
and subsequently summing up the cross sections of the $X^{1}\Sigma^{+}$ and
$B^{1}\Sigma^{+}$ states for the He$+$H$^{+}$ channel, and the cross sections of
the $A^{1}\Sigma^{+}$, $C^{1}\Sigma^{+}$ and $D^{1}\Sigma^{+}$ states for the
He$^{+}+$H channel.
In the N-SQR representation there are no electronic states to help us resolve
the two fragmentation channels. Nonetheless, they can be separated by projecting
the wavefunction onto the occupation-number sub-space in which all H-atom
spin-orbitals are empty with the projector
\begin{eqnarray}
    \mathcal{\hat{P}}_{H^{+}} = \prod_{i} (1-\hat{n}_{i_H}) ,
    \label{eq:proj_Hplus}
\end{eqnarray}
where $i_H$ is the index of the molecular spin-orbitals with pure
$H$ character.  These are well defined in the region of the CAP. This projector
singles out the part of the wavefunction dissociating into the He$+$H$^{+}$
channel, and \mbox{$\mathcal{\hat{P}}_{H} = 1- \mathcal{\hat{P}}_{H^{+}}$}
projects onto the remaining channel, He$^{+}+$H.

\begin{figure}[t]
    \begin{center}
    \includegraphics[width=8.5cm]{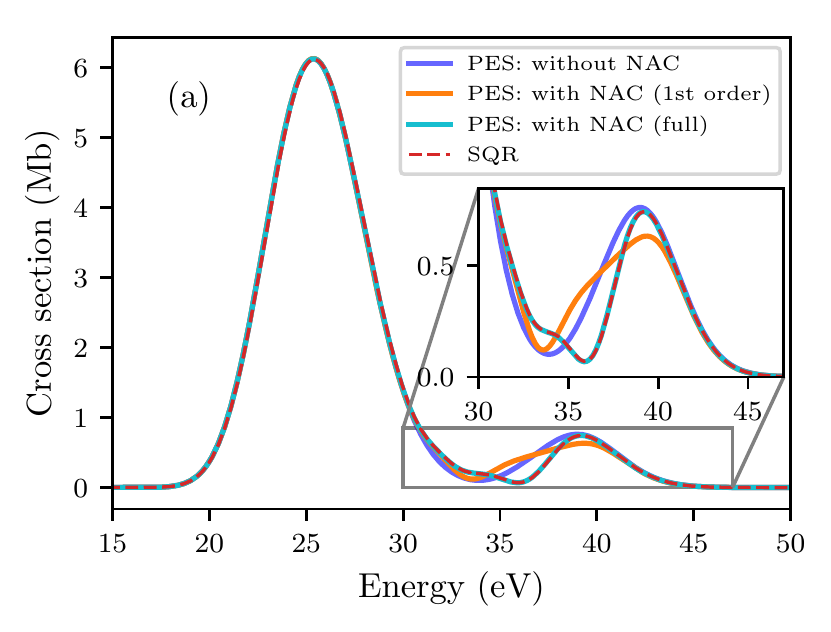}
    \includegraphics[width=8.5cm]{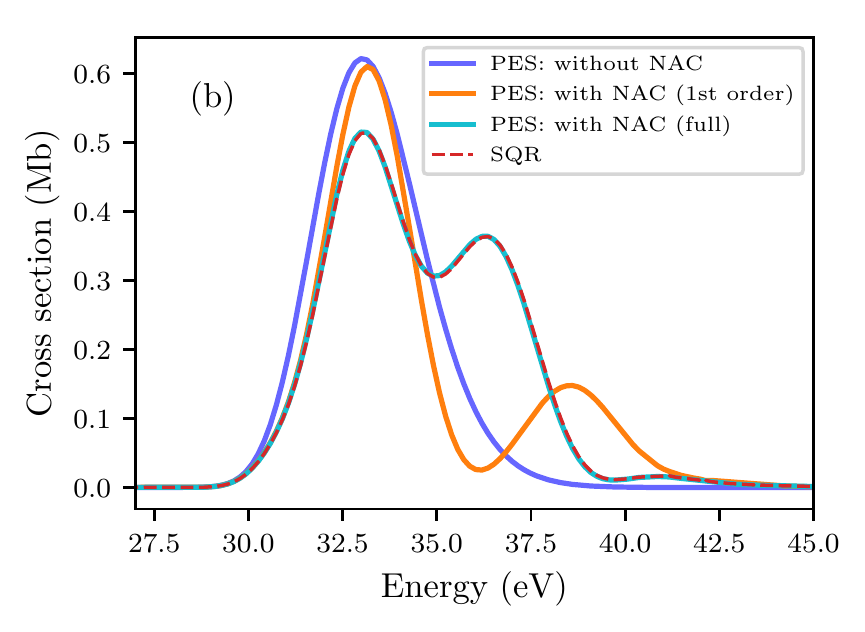}
    \caption{
        Photodisssociation cross section into
    (a) the He$^{+}$+H and
    (b) the He+H$^{+}$ channels.
    }
    \label{fig:flux_H}
    \end{center}
\end{figure}
Fig. \ref{fig:flux_H}a shows the photodissociation cross section of HeH$^{+}$
into He$^{+}$+H.
In the FQBH description the largest peak centered at about 25~eV
originates from dissociation in the $A^{1}\Sigma^{+}$
state. This state is energetically far from the next group of excited states
and there is negligible participation of it in non-adiabatic population transfer
between the dissociation channels.
The cross section in the region of this peak is identical for the FQBH
calculations with and without NAC, and N-SQR.

On the other hand, the peak centered between 38 and 40~eV arises due to
the dissociation through $C^{1}\Sigma^{+}$ and $D^{1}\Sigma^{+}$ electronic
states, which feature non-adiabatic couplings between them and with
the $B^{1}\Sigma^{+}$ electronic state.
The inset in Fig. \ref{fig:flux_H}a illustrates how the cross section in this
energy region is neither correctly described by an adiabatic description without
NACs nor by the inclusion of only first order NAC terms~\cite{nac1}. The FQBH with full NAC
and the N-SQR calculation (fully converged) agree exactly.
The importance of the second order NAC terms ($D_{pq}^\alpha$ in Eq.
\ref{eq:tdse_1st_sym}) is also highlighted here, as the calculation without the
second order contribution cannot reproduce the shape and the total area of the
spectrum.

The non-adiabatic effects in the photodissociation cross section of channel
He+H$^{+}$, shown Fig. \ref{fig:flux_H}b, are even more pronunced.
The adiabatic FQBH calculation presents a single peak at 38~eV, which correlates
with population initially transferred to the $B^{1}\Sigma^{+}$ electronic state
by the photoabsorption process.
A second peak at 36 to 38~eV appears due to the non-adiabatic transitions from
the $C^{1}\Sigma^{+}$ and $D^{1}\Sigma^{+}$ states to the $B^{1}\Sigma^{+}$
electronic state as the molecule dissociates.
Again, the calculation with only first order NAC terms~\cite{nac1} does not
reproduce the spectrum correctly; the second peak appears too high in
energy and the
intensity and the area are not correct, highlighting again the importance of
the second order terms.
On the other hand, the N-SQR calculation agrees exactly with the FQBH
calculation with the full first and second order NACs. This means, the
non-adiabatic transitions and interference effects between the photodissociation
channels are correctly captured within the N-SQR formulation. Moreover, one sees
how, in this particlar example, the simple diabatization of the molecular
orbitals based on rearranging them to guarantee the continuity of the
corresponding matrix elements is a very good approximation. All
relevant non-adiabatic effects are captured by the propagated N-SQR
wavefunction.
We emphasize again, that we could neglect the non-adiabatic coupling terms
Eqs.~(\ref{eq:ham_nuc_sqr}--\ref{eq:nac2_expression}) in the N-SQR calculations.
The $R$-dependence of the orbitals is rather weak, whereas the $R$-dependence of
the adiabatic state functions can be very strong.

\subsubsection{Convergence and numerical aspects of N-SQR}

\begin{figure}[t] \begin{center}
        \includegraphics[width=8.5cm]{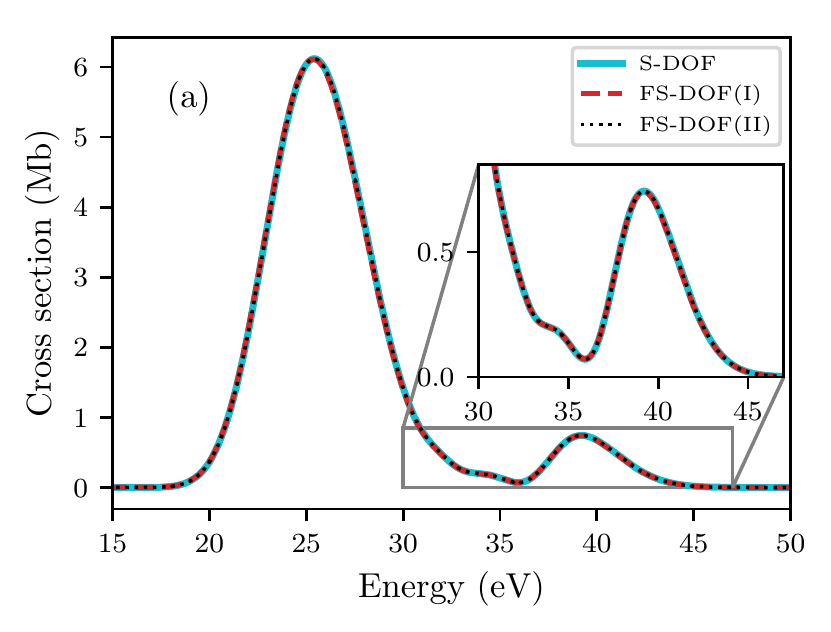}
        \includegraphics[width=8.5cm]{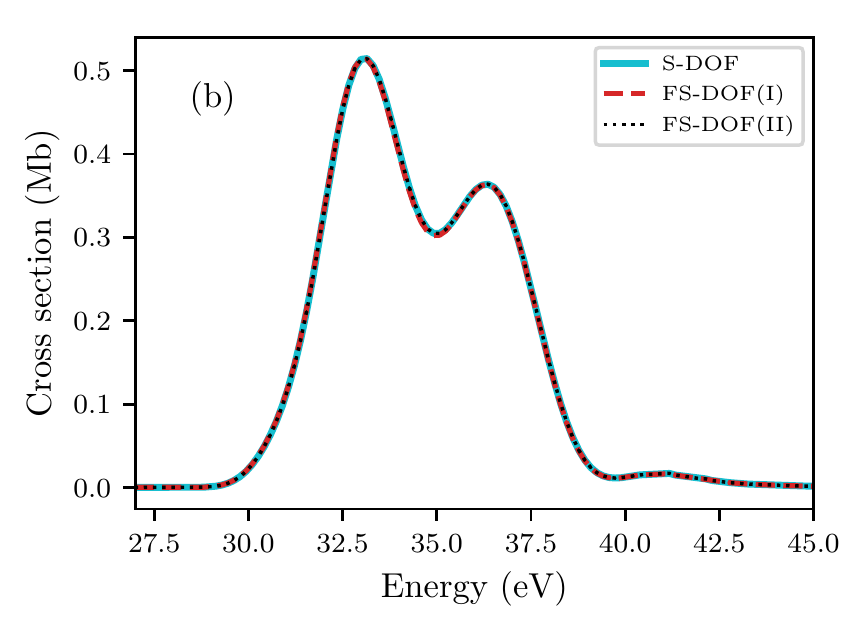}
        \caption{Photodissociation cross section of HeH$^+$ using the S-DOF and
        FS-DOF representations.  (a) He$^+$+H channel. (b) He+H$^+$ channel}
    \label{fig:flux_S_FS_DOF} \end{center} \end{figure}
\begin{figure}[t]
    \begin{center}
    \includegraphics[width=8.5cm]{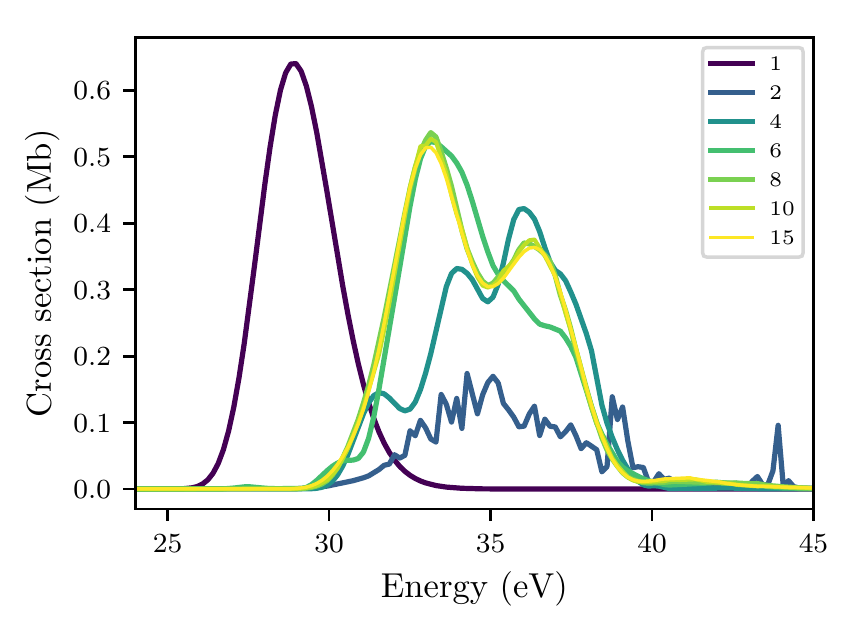}
    \caption{Photodissociation cross section of HeH$^+$ into He+H$^{+}$
        channel with different
        number of SPFs}
    \label{fig:spf_conv}
    \end{center}
    \end{figure}

    Figure~\ref{fig:flux_S_FS_DOF} compares the photodissociation cross section
    for the numerically converged S-DOF, FS-DOF(I) and FS-DOF(II) calculations
    and for the two dissociation channels.
    Table~\ref{tab:s_fs_dof} compares the number of SPFs, number of Hamiltonian
    terms and CPU time of the three different calculations, for which the
    corresponding wavefunction trees are shown in Fig.~\ref{fig:s_fs_dof_heh+}.
    When brought to numerical convergence, the various representations yield, as
    expected, identical results.
    %
    \begin{table}[t] 
    \caption{Number of Hamiltonian operator terms and
        wall-clock time for three N-SQR calculations.  The S-DOF and FS-DOF calculations
        use the tree in Fig.~\ref{fig:s_fs_dof_heh+}. The calculations have been
        performed with 16 CPUs using shared-memory parallelization on the same machine
        and CPU type, namely, Dual-Core Intel Xeon, processor type E5-2650 v2 running 
        at 2.6 GHz and the wall-clock times are intended for their relative comparison only.}
         \begin{ruledtabular}
          { \begin{center}
           \begin{tabular}{lccr}
            DOF repr.  & Hamil. terms & SPFs & Wall time (h:m) \\
            \hline
             S-DOF & 2100         & $(15, 9, 4, 9)$  & 64:18 \\ 
             FS-DOF(I)    & 1812  & $(15, 9, 4, 9)$  & 45:25 \\ 
             FS-DOF(II)   & 1300  & $(15, 15)$       & 11:15 \\ 
             FS-DOF(II)   & 1300  & $(10, 10)$       & 1:01 \\ 
             FS-DOF(II)   & 1300  & $(8, 8)$         & 0:15 \\ 
         \end{tabular}
          \end{center} }
          \end{ruledtabular}
    \label{tab:s_fs_dof} 
    \end{table}
    %
    Next, we focus on the FS-DOF(II) calculations and study their convergence
    behaviour. When the 10 spin-orbitals corresponding to a Fock-space of 1024
    states are combined and one uses that only one spin-$\alpha$ and one
    spin-$\beta$ electrons is present, this number reduces to $2\times
    \binom{5}{1}=25$ possible primitive states for the electrons with total spin
    projection onto the $z$-axis $S_z=0$.
    The nuclear DOF ($R$-grid) consists of 512 grid points and spans the range
    (0.5, 30.0) au.
    The photodissociation cross section of HeH$^{+}$ into He+H$^{+}$ channel is
    shown in Fig. \ref{fig:spf_conv} for an increasing number of SPFs for the
    electronic and nuclear modes, which are chosen equal to avoid redundant
    configurations.
    For the Hartree product case (SPF=1), the cross section features only one
    peak, which is centered in the wrong energy region. This is because of the
    complete neglect of correlation between electronic and nuclear DOF.
    Note that, since in this example the electronic spin-orbitals are fully
    combined into a single FS-DOF, the only approximation in the total
    wavefunction is the missing electron-nuclear correlation. The electronic
    correlation is treated, within the FS-DOF, exactly.
    The addition of a second SPF already brings the photodissociation cross
    section to the correct energy region centered around 35~eV. 6~SPFs provide
    already a qualitatively correct description, with a double-shoulder
    structure, and 8 SPfs and above yield the numerically converged result.

    It is interesting to note that, within the 25 primitive states with $S_z=0$
    in the FS-DOF it is possible to form 15 spin-singlet and 10 spin-triplet
    linear combinations, or spin-adapted configurations in the language of
    quantum chemistry. Therefore, a numerically exact calculation of the
    photodissociation cross section in the spin-singlet manifold requires 15
    SPFs. Convergence of the MCTDH-SQR wavefunction is reached for a smaller
    number of SPFs because some of the primitive states correspond to occupation
    patterns of the spin-orbitals with a very high energy and consequently a
    negligibly small population.

    \begin{figure}[t]
    \begin{center}
    \includegraphics[width=8.5cm]{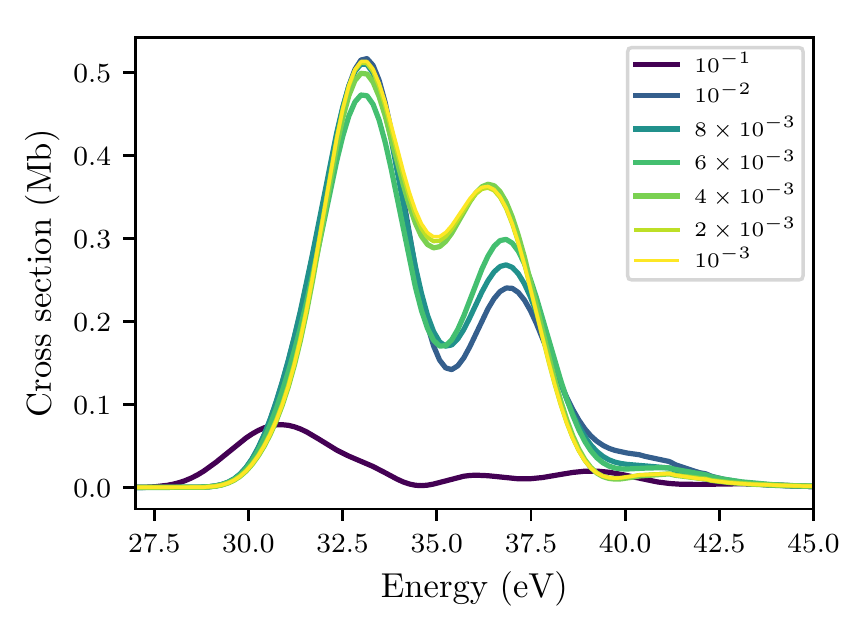}
    \caption{Comparison of photodissociation cross section of HeH$^+$ into
        He$+$H$^{+}$ channel with different cutoff for the 3- and 4- index terms
        of the Hamiltonian. Cutoff for 2-index terms is 10$^{-9}$ a.u.}
    \label{fig:cutoff_comp}
    \end{center}
    \end{figure}

    \begin{table}[t]
    \caption{Number of Hamiltonian terms with different cutoff values
       for the 3- and 4- index terms of the Hamiltonian}
    \begin{ruledtabular}
    {
    \begin{center}
    \begin{tabular}{lr}
    Cutoff   & Number of Hamiltonian terms \\ 
    \hline
    $10^{-1}$  & 140            \\ 
    $10^{-2}$  & 876            \\ 
    8$\cdot 10^{-3}$ & 956      \\ 
    6$\cdot 10^{-3}$ & 1068     \\ 
    4$\cdot 10^{-3}$ & 1204     \\ 
    2$\cdot 10^{-3}$ & 1284     \\ 
    $10^{-3}$  & 1300           \\ 
    $10^{-9}$  & 1300              
    \end{tabular}
    \end{center}
    }%
    \end{ruledtabular}
    \label{tab:cutoff_fs_dof}
    \end{table}
    Finally, we discuss the convergence of the photodissociation cross section
    with of the He$+$H$^{+}$ fragmentation channel as a function of the cutoff
    value for the Hamiltonian terms.
    The costs of an MCTDH calculation scale roughly linearly with the number of
    Hamiltonian terms, and therefore it is important to keep this number as low
    as possible.
    As was mentioned previously, we vary only the cutoff value for the 3- and
    4-index terms of the two-electron part of the electronic Hamiltonian,
    whereas the 2-index terms (Coulomb and exchange integrals) and the
    one-electron terms have in all cases a much smaller cutoff of 10$^{-9}$~au,
    which is standard in most MCTDH calculations~\cite{mctdh:MLpackage}.
    The cutoff values are given in Table~\ref{tab:cutoff_fs_dof}. The maximum
    number of terms, 1300, refers to the FS-DOF setting, which is already
    smaller than the maximum number of terms of the equivalent S-DOF
    calculation, 2100.
    A cutoff of $10^{-1}$ results in the cross section being qualitatively
    wrong, as seen in Fig.~\ref{fig:cutoff_comp}. However, the features are in
    the right energy region because the energy of the diagonal matrix elements
    between Fock states
    of the N-SQR Hamiltonian are not affected by the cutoff.
    A cutoff of $10^{-2}$ with roughly 70\% of the uncut Hamiltonian terms
    results in a qualitatively correct photodissociation, and convergence is
    reached with $4\cdot10^{-3}$~au, which includes about 80\% of the
    Hamiltonian terms.
    Although cutting Hamiltonian terms is not very helpful in the present case,
    it remains to be benchmarked in more detail in future works on larger
    molecules.

    \subsection{Properties of the N-SQR approach}
    \label{sec:properties}

    The N-SQR approach has some limitations. A major drawback is the large
    number of Hamiltonian terms resulting from the second-quantized description
    of the electrons, which grows proportionally to $M^4$, where $M$ is the
    number of spin-orbitals used to describe the electronic Fock space.
    This growth can be mitigated by introducing a cutoff for
    three- and four-index electronic integrals, by freezing orbital occupations,
    or by pruning configurations from FS-DOFs.
    (cf. Tab.~\ref{tab:s_fs_dof}). For example, we have
    pruned the FS-DOFs by removing occupation
    basis states that are incompatible with the total number of $\alpha$
    and $\beta$ electrons of the system.
    The FS-DOF representation has the added advantage over S-DOFs that the
    resulting operator terms acting on the FS-DOFs are extremely sparse.  This
    opens the possibility towards avoiding explicit matrix-vector
    multiplications in their application to the wavefunction.

    The other major drawback of the approach is the fact that the matrix
    elements of the second-quantized electronic Hamiltonian, the one- and
    two-body integrals $h_{ij}(\bm{R})$ and $v_{ijkl}(\bm{R})$, become
    nuclear-space potential operators, in general of dimensionality $3N-6$
    ($3N-5$ for linear systems), where $N$ is the number of atoms in the system.
    We have limited this first application of the N-SQR approach
    to a diatomic system, where the electronic integrals are 1-dimensional
    functions of the internuclear distance.
    Although applications to multidimensional systems have not been undertaken
    yet, there are grounds to believe that this will be feasible. One- and
    two-body integrals have a much simpler dependency on the nuclear coordinates
    than potential energy surfaces, and they can possibly be fit with simple
    functional forms and few parameters each.
    It may also be possible to expand the electronic integrals around a
    reference geometry up to, e.g., second order, thus yielding the equivalent
    of a vibronic coupling Hamiltonian in an N-SQR representation.

    A major feature of the N-SQR approach is that it circumvents the
    construction of potential energy surfaces before the quantum dynamics
    simulation, and hence it is not based on a group Born-Oppenheimer
    approximation~\cite{Born-Crystal,wor04:127}.
    This can be advantageous in situations where the number of relevant
    electronic states that would have to be pre-calculated becomes very large, thus
    complicating the quantum chemistry and diabatization procedures involved.
    Possible applications include, for example, reactions involving
    metal complexes, in which non-adiabatic effects are intermixed with the
    strong multi-configurational character of the electronic wavefunction
    %
    %
    ~\cite{bre09:489,nal14:44304}.
    In the photodissociation case we have studied, the observable is the
    fragmentation channel-resolved cross section. The adiabatic electronic
    states and potentials are not directly related to these quantities and are
    not observable in experiments that would determine the yield of each charged
    fragment only. In FQBH calculations, PES and NACs play often
    an auxiliary role to reach the actual measurable quantities.

    However, even though the group Born-Oppenheimer approximation is
    circumvented in N-SQR, the approach does not treat nuclei and electrons
    \emph{on the same footing}~\cite{ulu12:054112,hax11:63416}.
    The latter results in large couplings between
    the positions of the nuclei and the electrons due to their Coulomb
    attraction.
    Instead, the electrons appear just through the occupation of the
    spin-orbitals, from which only pre-computed one- and two-body integrals are
    needed.
    The underlying basis of molecular spin-orbitals can be generated through
    some previous first-quantization mean-field calculation and then properly
    diabatized, like in this work, or by some other suitable position-dependent
    transformation of atomic orbitals.
    The introduction of the electrons via the matrix elements of single-particle
    states instead of via the matrix elements of all-electron states (i.e. PES, NAC)
    provides great flexibility and, crucially, it removes the trivial
    ``\emph{electrons-follow-the-nuclei}'' correlation similarly as in the
    Born-Oppenheimer approximation.

    A clear advantage of circumventing the introduction of many-body
    electronic states is
    that only the molecular spin-orbitals need to be diabatized, which is a much
    simpler task than diabatizing electronic states~\cite{pac91:6668, dom93:362,
    gar97:161, nak01:10353, nak02:5576,yan13:84}.  Alternatively, the
    corresponding orbital non-adiabatic couplings in
    Hamiltonian~(\ref{eq:ham_nuc_sqr}) may be computed.
    As we have seen in the HeH$^+$ example, once diabatic orbitals and their
    matrix elements are introduced, the non-adiabatic dynamics is fully captured
    within the time-evolution of the nuclear-electronic wavepacket.
    %

\section{Summary and Conclusions}        \label{sec:conclusions}

    In this work we introduce an \emph{ab initio} quantum mechanical approach to
    treat the non-adiabatic quantum dynamics of electrons and nuclei based on a
    first quantization representation of the nuclei and a second quantization
    representation (SQR) of the electronic subsystem that overcomes the calculation of
    potential energy surfaces.
    We derive the full non-relativistic molecular Hamiltonian in this mixed
    representation, the N-SQR Hamiltonian~(\ref{eq:ham_nuc_sqr}), in which all
    degrees of freedom, nuclear coordinates and electronic spin-orbital
    occupations, are distinguishable, thus making the wavefunction
    representation amenable to the multi-configurational time-dependent Hartree
    (MCTDH) \emph{ansatz} and to its multilayer generalization, i.e. to
    a hierarchical Tucker tensor decomposition. Possibly,
    other related tensor contraction approaches can also be used.
    In this work we describe the application of the N-SQR Hamiltonian in
    conjunction with the MCTDH-SQR method for the time-propagation of the
    wavefunction, which, for fermions, is based on a Jordan-Wigner
    transformation of the fermionic operators to ladder spin-$1/2$ and auxiliary
    sign-change operators.

    After describing the general aspects of the MCTDH-SQR approach for fermions,
    we introduce the concept of Fock-space degrees of freedom (FS-DOF), which
    group together several spin degrees of freedom into a sub-Fock space.  This
    is shown to have several advantages: unneeded primitive occupation states
    can be pruned from the FS-DOF, which automatically reduces the number of
    Hamiltonian terms in the SQR. Operator terms acting on the
    FS-DOFs are very sparse, thus making their application more efficient.
    In this paper's application we have grouped up to 10 spin-orbitals in an
    FS-DOF, which results in 1024 internal
    states of the primitive degree of freedom before pruning, and
    larger FS-DOF can be handled by our code.

    The N-SQR approach requires either diabatic orbitals or the calculation of
    the orbital non-adiabatic couplings (NACs)
    in Hamiltonian~(\ref{eq:ham_nuc_sqr}). Here we have used
    a simple diabatization procedure based on the rearrangement of the molecular
    orbitals for diatomic systems and eventual phase corrections. With these
    orbitals, we obtain a perfect agreement between N-SQR and
    traditional first quantization Born-Huang (FQBH) calculations based on
    adiabatic electronic eigenstates and NAC elements between them.
    We also describe how to choose the initial SPFs for the corresponding MCTDH
    calculations, which in the context of MCTDH-SQR is, as recently pointed out
    by Manthe, of utmost importance.

    We apply the N-SQR approach to the photodissociation cross section of the
    HeH$^{+}$ molecule in the extreme ultraviolet photon energy range and focus
    on the parallel polarization case ($X^1\Sigma\to ^1\Sigma$ transitions),
    which has been studied theoretically~\cite{sae03:033409,dum09:165101} and
    experimentally (at the single photon energy of
    $38.7$~eV)~\cite{ped07:223202} before.
    HeH$^{+}$ fragments into either He+H$^+$ or He$^+$+H. The fragmentation
    channel-resolved photodissociation cross section in the energy region
    between 35 and 40~eV presents interference structures that originate from
    the electron transfer between the two atoms as they are starting to
    separate. These effects are particularly pronounced in the He+H$^+$ channel,
    where the highest energy peak of the cross section is
    absent in an adiabatic picture.
    There is complete agreement between the fully converged N-SQR and FQBH
    calculations including the full electronic state NACs up to second
    order in the adiabatic representation. Neglecting the second order NACs in
    the FQBH calculations does not lead either to the correct cross sections.

    We also study the accuracy of the N-SQR calculation as a function of the
    number of SPFs used in the MCTDH wavefunction.
    The calculations improve monotonically to the
    exact result with the number of SPFs, and by changing the number of SPFs of
    the various degrees of freedom the
    amount of correlation between nuclei and electrons, and between different
    sub-Fock spaces of the electronic system, can be controlled.
    Presently, we have implemented the mapping strategy for the
    application of mode operators to FS-DOF degrees of freedom within the
    Heidelberg MCTDH package. We still see room for improvements in the
    integration of the (ML-)MCTDH equations of motion in N-SQR applications. We
    emphasize that we are introducing the formalism and a proof-of-concept
    application, and that emphasis has not been put yet on computational
    performance.

    The main drawbacks of the \emph{ab initio} N-SQR approach have been
    presented and include the scaling of the number of Hamiltonian terms in
    the SQR with the 4th power of the number of spin-orbitals. We have discussed
    how these drawbacks can be, in part, mitigated.
    On the other hand, N-SQR circumvents the introduction of either adiabatic or
    diabatic electronic states, which can be of much advantage in situations
    where the number of electronic states that need to be considered in
    non-adiabatic processes is large and their energy spacings are
    commensurate with typical separations between vibrational energy levels.
    However,
    it is important to recognize that the N-SQR strategy to treat the coupled
    nuclear and electronic dynamics in molecules is not a silver bullet.
    Situations in which the quantum dynamics can be captured within a window of
    few electronic states will, in general, not benefit from the approach.
    Through the selected example we illustrate the salient features of the method,
    hint at possible application domains,
    and lay down the theoretical and practical foundations for subsequent
    developments and approximations based on the N-SQR formalism.
    The actual range of applications where the approach will prove to be useful
    remains an open question to be explored in future work.

\section{Supplementary Material}

    See the supplementary material for the tabular data of the
    NACME, dipole moment and transition dipole moments among the ground and
    excited electronic states of HeH$^+$.

\section{Data Availability}
    The data that support the findings of this study are available in tabular
    form in the supplementary materials.
\section{Acknowledgments}

    We thank Prof. H.-D. Meyer for his comments on the manuscript and
    important assistance with the MCTDH calculations.
    The authors declare no conflicts of interest.


%

\appendix

 \section{Born-Huang expansion with the molecular SQR Hamiltonian}
    \label{ap:bh}

    As mentioned in the main text, Hamiltonian~(\ref{eq:ham_nuc_sqr}) is
    compatible with low-rank tensor decompositions of the wavefunction owing to the
    distinguishability of the degrees of freedom.
    Nonetheless, it is illustrative to consider formally a wavefunction
    expansion analogous to the usual Born-Huang (BH) \emph{ansatz}, in which all
    occupation-number states are explicitly kept:
    \begin{align}
        \label{eq:ansatz}
        |\Psi (t)\rangle = \sum_{\{\vec{n}\}}
        \Phi_{\{\vec{n}\}}(\mathbf{R},t)
        |{\{\vec{n}\}}\rangle
    \end{align}
    where $|{\{\vec{n}\}}\rangle$ is the occupation number representation
    \textit{ket} of an
    electronic configuration uniquely described by the set of spin-orbital
    occupations $\vec{n}$ in terms of the
    $\psi(\vec{x},\mathbf{R})$ basis
    and $\Phi_{\{\vec{n}\}}(\mathbf{R},t)$ is the corresponding probability
    amplitude, which, as in the electronic states formulation, is a
    function of the nuclear positions.

    Inserting the \emph{ansatz}~(\ref{eq:ansatz}) into the general TDSE while using
    Hamiltonian~(\ref{eq:ham_nuc_sqr}) results in the TDSE
    \begin{widetext}
    \begin{align}
        \label{eq:tdse4}
        i\dot{\Phi}_{\{\vec{m}\}}(\mathbf{R},t) & =
        \left[
        \hat{T}_n
        +
        \hat{W}_n
        \right]
        \Phi_{\{\vec{m}\}}(\mathbf{R},t)
        \\ \nonumber
        & +
        \sum_{\{\vec{n}\}}
        \left(
        \sum_{pq}^F
            \left[
                {h}_{pq} (\mathbf{R})
                -\frac{1}{M}\mathbf{D}_{pq}(\mathbf{R})\cdot\boldsymbol{\nabla}
                -\frac{1}{2 M} G_{pq}(\mathbf{R})
            \right]
        \langle{\{\vec{m}\}}|
        \hat{b}_p^{\dagger} \hat{b}_q
        |{\{\vec{n}\}}\rangle
        \right.
        \\ \nonumber
        & +
        \left.
        \frac{1}{2}\sum_{pqrs}^F {v}_{pqrs}(\mathbf{R})
        \langle{\{\vec{m}\}}|
        \hat{b}_p^{\dagger} \hat{b}_q^\dagger
        \hat{b}_r \hat{b}_s
        |{\{\vec{n}\}}\rangle
        \right)
        \Phi_{\{\vec{n}\}}(\mathbf{R},t).
    \end{align}
    \end{widetext}
    In this BH-SQR formulation of the non-adiabatic dynamics problem, the TDSE
    contains both differential and potential coupling terms between the nuclear
    and electronic spaces.
    If the kinetic coupling terms are kept small or negligible by choosing a
    quasi-diabatic spin-orbital basis, the remaining non-adiabatic coupling
    effects are described directly by the time-evolution of the nuclear
    amplitudes $\Phi_{\{\vec{m}\}}(\mathbf{R},t)$.
    As usual, there is no free lunch and the price to be paid is a potentially
    much larger space of electronic occupation states as compared to adiabatic
    electronic states.
    Therefore, the BH variant of the nuclear-SQR approach, Eq.~(\ref{eq:tdse4}),
    may not find much direct use.
    Conceptually, it is completely equivalent to a first quantization description in
    which each electronic configuration of the full configuration-interaction
    matrix carries its own nuclear amplitude. For this reason,
    practical applications will still be based on the N-SQR formulation of the
    main text.

\section{Non-adiabatic quantum dynamics in the adiabatic representation}
\label{ap:fq}

The electronic part of the total molecular Hamiltonian (\ref{eq:hamTot1}) in the first quantization
representation is given as
\begin{eqnarray}
 \hat{H}_e = \sum_{i=1}^N \Big [ -\frac{1}{2}\nabla_i^2 - \sum_{\alpha=1}^M \frac{Q_\alpha}{r_{i\alpha}} +
 \sum_{j>i} \frac{1}{r_{ij}} \Big ]
\end{eqnarray}
Using the Born-Huang expansion, the total molecular wavefunction is expressed as
\begin{eqnarray}
 \Psi(x,R,t) = \sum_p \chi_p(x;R)\Phi_p(R,t)
\end{eqnarray}
where $\chi_n(x;R)$ are the electronic basis functions that parametrically depend on the nuclear coordinates and
$\Phi_n(R,t)$ are the nuclear wavefunction at time $t$. Now inserting this wavefunction ansatz into the time dependent
Schor\"{o}dinger equation, one obtains the following coupled equation
\begin{widetext}
\begin{eqnarray}
 i \dot{\Phi}_q(R,t) = \hat{T}_n \Phi_q(R,t)+\sum_p \left[ V_{pq}(R) -
 \sum_{\alpha} \frac{1}{2m_{\alpha}} \left( 2 d_{pq}^{\alpha}(R) \frac{\partial}{\partial R_{\alpha}} + G_{pq}(R) \right) \right] \Phi_p(R,t)
\label{eq:tdse_1st}
\end{eqnarray}
\end{widetext}
where
\begin{eqnarray}
 V_{pq}(R)=\langle \chi_p(x;R)|\hat{H}_e+\hat{W_n}|\chi_q(x;R)\rangle \\
 d^{\alpha}_{pq}(R) = \langle \chi_p(x;R)| \frac{\partial}{\partial R_\alpha} |\chi_q(x;R)\rangle \\
 G^{\alpha}_{pq} = \langle \chi_p(x;R)|\frac{\partial^2}{\partial R_{\alpha}^2} | \chi_q(x;R)\rangle
\end{eqnarray}
In the adiabatic representation, the electronic basis functions become the eigen functions of the electronic
Hamiltonian and thus, the $V$ becomes a diagonal matrix with adiabatic electronic energies in the diagonal.
Thus, for one nuclear coordinate, the non-adiabatic coupling operator between $i$th and $j$th electronic states  is
\begin{eqnarray}
 \Lambda_{ij} = \frac{1}{2M} \left( 2 d_{ij} \frac{\partial}{\partial R} + G_{ij} \right)
\end{eqnarray}
Now the first order term itself is not Hermitian. But we can symmetrize it by following 
\begin{eqnarray}
\frac{\partial}{\partial R} d_{ij} &=& \frac{\partial}{\partial R} \langle \phi_i | \frac{\partial}{\partial R} | \phi_j \rangle \nonumber \\
&=& \langle \frac{\partial \phi_i}{\partial R}  |  \frac{\partial \phi_j}{\partial R} \rangle
+ \langle \phi_i | \frac{\partial^2}{\partial R^2} | \phi_j \rangle
+ \langle \phi_i | \frac{\partial}{\partial R} | \phi_j \rangle \frac{\partial}{\partial R} \nonumber \\
&=& \langle \frac{\partial \phi_i}{\partial R}  |  \frac{\partial \phi_j}{\partial R} \rangle
+ G_{ij}
+ d_{ij} \frac{\partial}{\partial R} \nonumber \\
&=& \sum_l \langle \frac{\partial \phi_i}{\partial R}  | \phi_l \rangle \langle \phi_l |  \frac{\partial \phi_j}{\partial R} \rangle 
+ G_{ij}
+ d_{ij} \frac{\partial}{\partial R} \nonumber \\
&=& \sum_l d_{il}^{*}d_{lj} + G_{ij} + d_{ij} \frac{\partial}{\partial R}
\end{eqnarray}
Adding $d_{ij}\frac{\partial}{\partial R}$ to both sides and rearranging
\begin{eqnarray}
2 d_{ij} \frac{\partial}{\partial R} + G_{ij} = \frac{\partial}{\partial R} d_{ij} + d_{ij} \frac{\partial}{\partial R} - D_{ij}
\end{eqnarray}
where
\begin{eqnarray}
D_{ij} = \sum_l d_{il}^{*}d_{lj}.
\label{nac_order2}
\end{eqnarray}
So, the symmetrized form of the NAC operator reads
\begin{eqnarray}
\Lambda_{ij} = \frac{1}{2M} \left( \frac{\partial}{\partial R} d_{ij} + d_{ij} \frac{\partial}{\partial R} - D_{ij} \right) .
\end{eqnarray}
Using this symmetrized form of the coupling terms, the Eq. \ref{eq:tdse_1st} is written as
\begin{widetext}
\begin{eqnarray}
 i \dot{\Phi}_q(R,t) = \hat{T}_n \Phi_q(R,t)+\sum_p \left[ V_{pq}(R) -
 \sum_{\alpha} \frac{1}{2m_{\alpha}} \left( \frac{\partial}{\partial R} d^{\alpha}_{pq}(R) + d^{\alpha}_{pq}(R) \frac{\partial}{\partial R} - D^{\alpha}_{pq}(R) \right) \right] \Phi_p(R,t) .
\label{eq:tdse_1st_sym}
\end{eqnarray}
\end{widetext}

To study the nonadiabatic effects in the photodissociation of HeH$^+$,
the uncontracted aug-cc-pVTZ basis is used for both H and He atoms.
In this work we have considered
10 lowest energy HF spin orbitals of $\Sigma$ symmetry
to calculate the potential energy curves, non-adiabatic coupling matrix
elements and the transition dipole moments between the lowest 5
$^1\Sigma^{+}$ states.
Those are given in the supplementary material.

\end{document}